\documentclass[a4paper,11pt]{article}
\usepackage{lineno}

\pdfoutput=1 

\usepackage{jheppub} 

\usepackage[T1]{fontenc}

\usepackage{graphicx} 
\graphicspath{{figures/}} 
\usepackage{subcaption}   

\usepackage{multirow}

\usepackage{placeins}

\title{\boldmath Searching for anomalous top quark interactions with proton tagging and timing detectors at the LHC}

\author[a]{Cristian Baldenegro,}
\author[b]{Andrea Bellora,}
\author[c,d]{Sylvain Fichet,}
\author[e]{Gero von Gersdorff,}
\author[f]{Michael Pitt,}
\author[g]{Christophe Royon}

\affiliation[a]{\'{E}cole Polytechnique, Laboratoire Leprince-Ringuet, Av. Chasles, 91120 Palaiseau, France}
\affiliation[b]{INFN Sezione di Torino and Università degli Studi di Torino, Via P. Giuria 1, 10125 Torino, Italy}
\affiliation[c]{ICTP South American Institute for Fundamental Research  \& IFT-UNESP,
R. Dr. Bento Teobaldo Ferraz 271, S\~ao Paulo, Brazil
}
\affiliation[d]{
Centro de Ciencias Naturais e Humanas, Universidade Federal do ABC, Santo Andre, 09210-580 SP, Brazil
}
\affiliation[e]{Departamento de Fısica, Pontifıcia Universidade Catolica de Rio de Janeiro, Rio de Janeiro, Brazil}
\affiliation[f]{CERN, CH\- 1211, Geneva 23, Switzerland}

\affiliation[g]{The University of Kansas, Department of Physics and Astronomy, 1251 Wescoe Hall Dr , 66045 Lawrence, Kansas, US}

\newcommand{\be}{\begin{equation}}
\newcommand{\ee}{\end{equation}}

\abstract{

We study the LHC sensitivity to new broad neutral resonances produced in two-photon fusion and decaying to a top quark pair, $\gamma\gamma \to t\bar{t}$. This is probed in central exclusive $t\bar{t}$ production in proton-proton collisions, $pp \to p t\bar{t} p$. We use the tagging of the intact protons by PPS (CMS) and AFP (ATLAS)  and consider the semi-leptonic $t\bar t$ channel. The sensitivity is also mapped onto a set of dimension-8 $\gamma\gamma t\bar{t}$ operators in the large mass limit. 
Using the kinematical correlations between the intact protons and the reconstructed $t\bar{t}$ system, we obtain a sensitivity to the couplings of the dimension-8 operators of $1.4 \cdot 10^{-11}$ GeV$^{-4}$ at 95\%\,CL. The sensitivity to the anomalous couplings is significantly improved down to about $7\cdot 10^{-12}$ GeV$^{-4}$ if the proton time-of-flight is known with a precision of 20\,ps in future measurements.
The 95\%\,CL sensitivity to broad neutral resonances  reaches masses of order $ 1500$\,GeV when using timing information. 

}

\begin{document}

\maketitle
\flushbottom

\section{Introduction}
\label{sec:intro}

Under the hypothesis that new physics with strong interactions exists beyond the Standard Model (SM) of particle physics,  it is possible that a new state with broad width lies at the TeV scale. 
Searching for such broad resonances at the LHC is somewhat challenging since the analyses cannot efficiently rely on standard bump-searching methods. Instead, the search  may substantially benefit from observation channels featuring a highly reduced background. The observation channels from central exclusive production (CEP) processes, in which both protons remain intact in the final state, belong to this category and therefore provide an environment favorable to the search for broad resonances. 

Among all processes, the CEP topology stands out because it can be efficiently selected using the forward proton detectors built for this purpose in the CMS and ATLAS experiments, namely the Precision Proton Spectrometer (PPS) \cite{PPStdr} and the ATLAS Forward Proton  detector (AFP) \cite{AFPtdr}, respectively. Forward detectors measure the outgoing intact protons, leading to the reconstruction of the entire final state kinematics of both the proton and central systems. 
This set of techniques leads to drastic background reduction. A key example is the  $pp \to\gamma\gamma pp$ process for which the background for $300$\,fb$^{-1}$ drops below one expected event in the high mass region (see \textit{e.g.} \cite{Fichet:2014uka}). It was demonstrated in \cite{Baldenegro:2018hng} that this channel is an efficient precision probe of broad neutral particles~\footnote{
Other studies of new physics searches based on proton tagging at the LHC 
  can be found in 
\cite{usww, usw,Sahin:2009gq,Atag:2010bh, Gupta:2011be, Epele:2012jn, Lebiedowicz:2013fta, Fichet:2013ola,Fichet:2013gsa,Sun:2014qoa,
Sun:2014qba,Sun:2014ppa,Sahin:2014dua,Inan:2014mua,
Fichet:2014uka,Fichet:2015nia,Cho:2015dha,Fichet:2016clq,Fichet:2015vvy,Fichet:2016pvq,
Baldenegro:2017aen,
Baldenegro:2017lzv}.
}.

The SM central exclusive production of $t\bar{t}$ has been investigated in recent phenomenological studies \cite{Luszczak:2018dfi, Goncalves:2020saa,Martins:2022dfg}. 
The process has not been observed experimentally, but a recent search by the CMS Collaboration has set an upper bound on its cross section of $\sigma = 0.59$ pb at 95\% CL at 13 TeV~\cite{CMS-PAS-TOP-21-007}. It is likely that the SM process could be observed at the high-luminosity LHC \cite{CMS:2021ncv}. Photoproduction of top quark pairs can be used to constrain the electromagnetic dipole moments of the top quark, as discussed in Ref. ~\cite{Koksal:2019gyo} for $\gamma$-gluon fusion processes, for example.

The goal of the present paper is to expand the prospects of CEP searches to the case of neutral particles coupled to both photons and top quarks. 
Our focus is on evaluating the sensitivity of the exclusive $pp\to t \bar t p p$ process to the presence of (possibly broad) neutral resonances. 
In the large mass limit the photon-top quark interactions become local and can be described using local dimension-8 effective operators. Thus, the sensitivity to these dimension-8 operators  is also studied throughout this paper as a natural extension. 

We emphasize that our focus on dimension-8 operators is motivated by the neutral particle scenario. In principle, searching for the dimension-6 top quark dipole operators in the CEP topology is also well-motivated. This interesting case is kept for a separate investigation. In the case of neutral particles, we will verify that the loop-generated contribution to the dipole is negligible in the scope of our study, hence justifying our focus on the dimension-8 operators.

Importantly,  the sensitivity to the $pp\to t \bar t  p p $ process is expected to be substantially enhanced by the use of timing detectors embedded into the forward detectors \cite{Cerny:2020rvp}.
These timing detectors provide an independent reconstruction of the primary vertex using intact proton information, improving further the selection of signal events. 
Our study  serves to illustrate the power of these timing detectors in the specific case of  broad neutral resonances.

The plan of the paper is as follows.
In Section \ref{se:theory} the models for neutral particles are introduced. We also work out the basis of effective operators describing local $\gamma\gamma t \bar{t}$ interactions.
 Section~\ref{se:simulation} provides basics on forward detectors and describes the simulation framework.  Section~\ref{se:stats} briefly describes  the statistical framework. 
 Section~\ref{se:lchannel} describes the search for $\gamma\gamma\to t\bar t$ in the semi-leptonic decay channel. 
 Section~\ref{se:results} presents our results and a summary is given in Section~\ref{se:summary}. 

\section{Effective field theory and neutral particles}
\label{se:theory}

Here, we define extensions of the SM which induce anomalous $\gamma\gamma\to t\bar t$ scattering. We start by deriving the relevant basis of higher dimensional operators that contribute to such anomalous processes. The lowest dimensional operators turn out to have dimension eight, and we identify three CP-even and three CP-odd operators. Any new physics model contributing to anomalous $t\bar t$ production via photon fusion can be matched to these operators at low energy. We then introduce generic models of neutral resonances with trilinear couplings to both photons and top quarks. They can themselves be considered subsectors of well-motivated extensions of the SM, such as composite Higgs models, extra dimensions, etc. Moreover, we explicitly provide the matching of these generic models to the aforementioned dimension-8 effective theory. These (effective) models provide the basis for our simulations in the subsequent sections.

\subsection{Operator basis}

The physics candidates probed by our experiment  may have a mass scale $\Lambda$ much higher than the other scales involved in the amplitudes. In such a regime,  the amplitudes can be expanded in inverse powers of $\Lambda$. Whenever this expansion is possible,  the observable  effects of the  high mass physics  can  be described by a series of local operators encoded in a low-energy effective Lagrangian, ${\cal L}={\cal L}_{\rm SM}+\sum_{n,i} \frac{a_{i,n}}{\Lambda^n}{\cal O}_{i,n}$\,. 

Our interest is in the set of dimension-8 operator (\textit{i.e.} $n=4$) involving two top quarks and two photons. 
 We allow for both CP-even and CP-odd bilinear invariant in $\bar t t$ and photon field strength. At a given order in the expansion, the equations of motion can be used to reduce the set of the effective operators. We find an irreducible basis of six operators (omitting the subscript $n$),
\begin{align}
    \mathcal{O}_1  &=  m_t F^{\mu\nu} F_{\mu\nu} \bar t t & 
    \mathcal{O}_2 &= i m_t F^{\mu\nu}\tilde{F}_{\mu\nu} \bar t \gamma_5 t
    \nonumber
    \\
        \mathcal{O}_3 &= m_t F^{\mu\nu} \tilde{F}_{\mu\nu} \bar t t  & 
 \mathcal{O}_4 &= i m_t F^{\mu\nu} F_{\mu\nu} \bar{t} \gamma_5 t  
 \nonumber
    \\
        \mathcal{O}_5 &= i  F^{\mu\rho}F^\nu_\rho \bar{t} \gamma_\mu D_\nu t  & 
            \mathcal{O}_6 &=  F^{\mu\rho}F^\nu_\rho \bar{t} \gamma_5 \gamma_\mu D_\nu t
            \label{eq:EFTbasis}
\end{align}
with $\tilde F_{\mu\nu}=\frac{1}{2}\epsilon_{\mu\nu\rho\sigma}F^{\rho\sigma}$. 
The ${\cal O}_{1\ldots 4}$ operators involve two derivatives, the ${\cal O}_{5,6}$ involve three derivatives. The ${\cal O}_{1,3,5}$ operators involve CP-even bilinear terms while  ${\cal O}_{2,4,6}$ involve CP-odd bilinear ones. 

Bilinear terms of the form $\bar{t} \Gamma \gamma_\mu \partial^\mu t$, where $\Gamma$ is an arbitrary Lorentz structure,  are reduced by using the top quark equation of motion into other operators from the above basis plus operators involving other fields, which are irrelevant for the process considered. One can also check that $F^{\mu\rho} \tilde{F}^\nu_\rho = \frac{1}{4} g_{\mu\nu} F \tilde{F}$ such that operators of the form $F^{\mu\rho} \tilde{F}^\nu_\rho \bar{t} \Gamma \gamma_\mu \partial_\nu t$ are reduced into $\mathcal{O}_{3,4}$ using the top quark equation of motion.

The effective Lagrangian description breaks down at momenta of order of the $\Lambda$ mass scale. In case of strong coupling, this breakdown coincides with the breakdown of the perturbative expansion in the effective couplings. It also roughly coincides with violation of unitarity, signalling the necessity of  a UV completion.

We parametrize the coefficients of the basis of operators shown in Eq.\,\eqref{eq:EFTbasis} as
\be
{\cal L}={\cal L}_{\rm SM}+\sum_{i=1}^6 \zeta_i {\cal O}_{i} \,. 
\ee

\subsection{Neutral particles}

Neutral particles with non-renormalizable couplings to SM operators are present in common extensions of the SM. Such theories often contain scalar, pseudo-scalar and spin-2 particles, respectively denoted by $\varphi$, $\tilde \varphi$ and $h^{\mu\nu}$. 
Examples include the Peccei-Quinn axion \cite{Peccei:1977hh}, the Kaluza-Klein (KK) graviton and the radion in warped extra dimensions~\cite{Randall:1999ee}, the dilaton in theories of strongly coupled electroweak breaking \cite{Goldberger:2007zk}, Goldstone bosons of extended composite Higgs models \cite{Kaplan:1983sm}, mesons and glueballs of strongly-coupled theories \cite{Hill:2002ap}, extra scalars breaking the global symmetry of composite Higgs models \cite{vonGersdorff:2015fta,Fichet:2016xvs,Fichet:2016xpw}, Higgs  portal models \cite{Schabinger:2005ei},
and  many more.

Independently of the model they originate from, their leading couplings to photons and quarks can be can be classified in terms of CP quantum numbers and written as  
\be
{\cal L }^{\gamma\gamma}_{\rm eff} = 
\frac{1}{f^{\gamma\gamma}_{{\bf 0}^+ }}\varphi (F_{\mu\nu})^2 + \frac{1}{f^{\gamma\gamma}_{{\bf 0}^- }} a F_{\mu\nu}\tilde F_{\mu\nu} + \frac{1}{f^{\gamma\gamma}_{{\bf 2}}} h^{\mu\nu}  \left(
-F_{\mu\rho} F^{\rho}_{\,\,\nu}
+ \frac{1}{4}\eta_{\mu\nu}(F_{\rho\sigma})^2\right) \,, 
\label{eq:resgg}
\ee
\be
{\cal L }^{\bar t t}_{\rm eff} = 
\frac{m_t}{f^{\bar tt}_{{\bf 0}^+ }} \varphi  \bar t t  +  i \frac{m_t}{f^{\bar tt}_{{\bf 0}^- }} a \bar t \gamma_5 t + i \frac{1}{f^{\bar tt}_{{\bf 2} }} h^{\mu\nu} \bar t \gamma_\mu D_\nu t  \,
\label{eq:restt}
\ee
where $\varphi$ is the CP-even scalar, $a$ the CP-odd scalar, and $h^{\mu\nu}$ the CP-even spin-2 field. 

The couplings $f_X$ appearing in Eq.(\ref{eq:resgg}) e (\ref{eq:restt}) are typically restricted by unitarity to satisfy 
\be
\frac{1}{f_X}\lesssim \frac{4\pi}{m}\,.
\label{eq:strong}
\ee
Let us note that in some models that address both the hierarchy and flavor problems of the SM (such as warped extra-dimensions with SM fermions in the bulk), it is common that the heavy neutral resonances couple more strongly to the top quark
than to the lighter quarks.

    The partial width of both the CP-even and CP-odd scalar into $\gamma\gamma$  is 
\be
\Gamma_{\phi_\pm \to \gamma\gamma}=\frac{m^3}{4\pi f^2_{\gamma\gamma}}\,.
\ee
The partial width for the CP-even and CP-odd scalar into $t\bar t$ are respectively 
\be
\Gamma_{\phi_+\to\bar t t } = \frac{3}{8\pi}\frac{m_t^2}{(f^+_t)^2} m \left(1-\frac{4m_t^2}{m^2}\right)^{3/2}
\ee
\be
\Gamma_{\phi_-\to\bar t t } = \frac{3}{8\pi}\frac{m_t^2}{(f^-_t)^2} m \left(1-\frac{4m_t^2}{m^2}\right)^{1/2}\,.
\ee

\subsection*{Matching to dimension-8 operators at tree-level} 

When any of these particles is heavy, it contributes to the effective EFT operators classified in Eq.\,\eqref{eq:EFTbasis}. The matching is as follows, depending on which of the particles is integrated out
\begin{itemize}
\item 
Spin-0, CP-even scalar $\phi$
\be
\zeta_1 = \frac{1}{f^{\gamma\gamma}_{{\bf 0}^+ }f^{\bar t t}_{{\bf 0}^+ }m^2 },\quad\quad\quad\quad\quad { \zeta_{i\neq 1}=0 }
\ee
\item 
Spin-0 CP-odd scalar $\tilde \phi$
\be
\zeta_2 = \frac{1}{f^{\gamma\gamma}_{{\bf 0}^- }f^{\bar t t}_{{\bf 0}^- }m^2 },\quad\quad\quad\quad\quad { \zeta_{i\neq 2}=0 }
\ee
\item
 Spin-2 field $h_{\mu\nu}$
\be
\zeta_1 = \frac{1}{4 f^{\gamma\gamma}_{{\bf 2} }f^{\bar t t}_{{\bf 2} }m^2 },\quad\quad\quad\quad\quad
\zeta_5 = - \frac{1}{ f^{\gamma\gamma}_{{\bf 2} }f^{\bar t t}_{{\bf 2} }m^2 },\quad\quad\quad\quad\quad
{\zeta_{i\neq 1,5}=0 }
\ee
\end{itemize}
This EFT regime typically applies in our context whenever $m\gtrsim 2.8$ TeV as the experimental acceptance is very low beyond this scale and a heavier resonance cannot be resolved, leaving only the  imprint in the higher dimensional operators.

\subsection*{Estimate of dipole operator at loop-level} 

The above couplings of the top quark to the neutral resonance also generate a CP-even dimension-6 dipole operator 
\begin{equation}
 \zeta_d   m_t F_{\mu\nu} \bar t\gamma^\mu\gamma^\nu\gamma^5t
\end{equation}
where a factor of $m_t$ arising from electroweak symmetry breaking is shown explicitly and $\zeta_d$ has a mass dimension of  $m^{-2}$.
This operator modifies the SM top quark coupling to the photon, and hence the $\gamma\gamma\to t\bar t$ production amplitude via the standard $t$-channel topology.
However, the coefficient $\zeta_d$ is loop suppressed and goes as
\begin{equation}
    \zeta_d\sim\frac{e}{16\pi^2}\frac{m_t^2}{(f^{tt})^2m^2}
\end{equation}
and even in the strong-coupling scenario in Eq.~(\ref{eq:strong}), $\zeta_d\to \frac{e m_t^2}{m^4}$  suffers an additional strong $ (\frac{m_t}{m})^2$ suppression. The combined effect for the modification of the SM coupling is
\be
\frac{\delta e}{e}\sim \frac{m_t^3 E}{m^4}
\ee
which is of the order of $\sim 10^{-6}$ for a 1 TeV resonance and a typical energy of $E=$500 GeV. Therefore, such an effect is completely negligible.

\section{Simulation framework}
\label{se:simulation}
\subsection{Signal modeling}
\label{sec:signal_simulation}

\begin{figure}[t]
     \centering
     \begin{subfigure}[b]{0.31\textwidth}
         \centering
         \includegraphics[width=\textwidth]{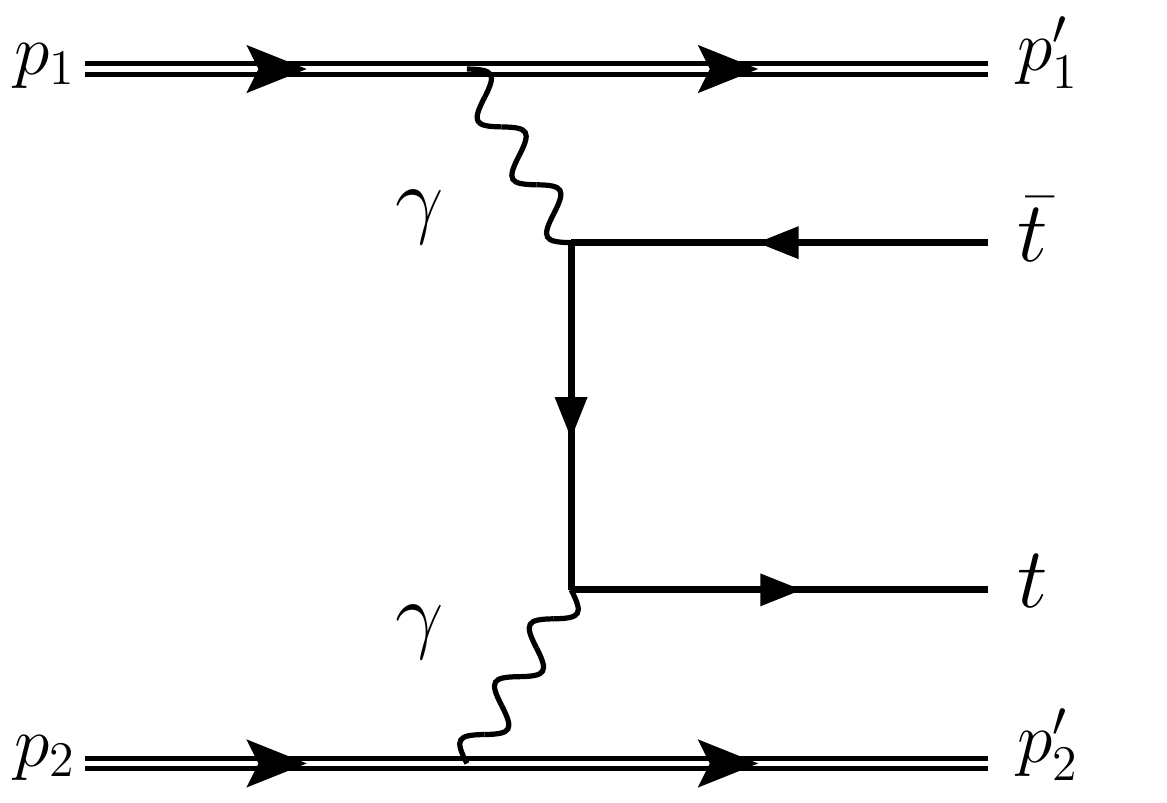}
         \caption{}
         \label{fig:SM_QED}
     \end{subfigure}
     \begin{subfigure}[b]{0.31\textwidth}
         \centering
         \includegraphics[width=\textwidth]{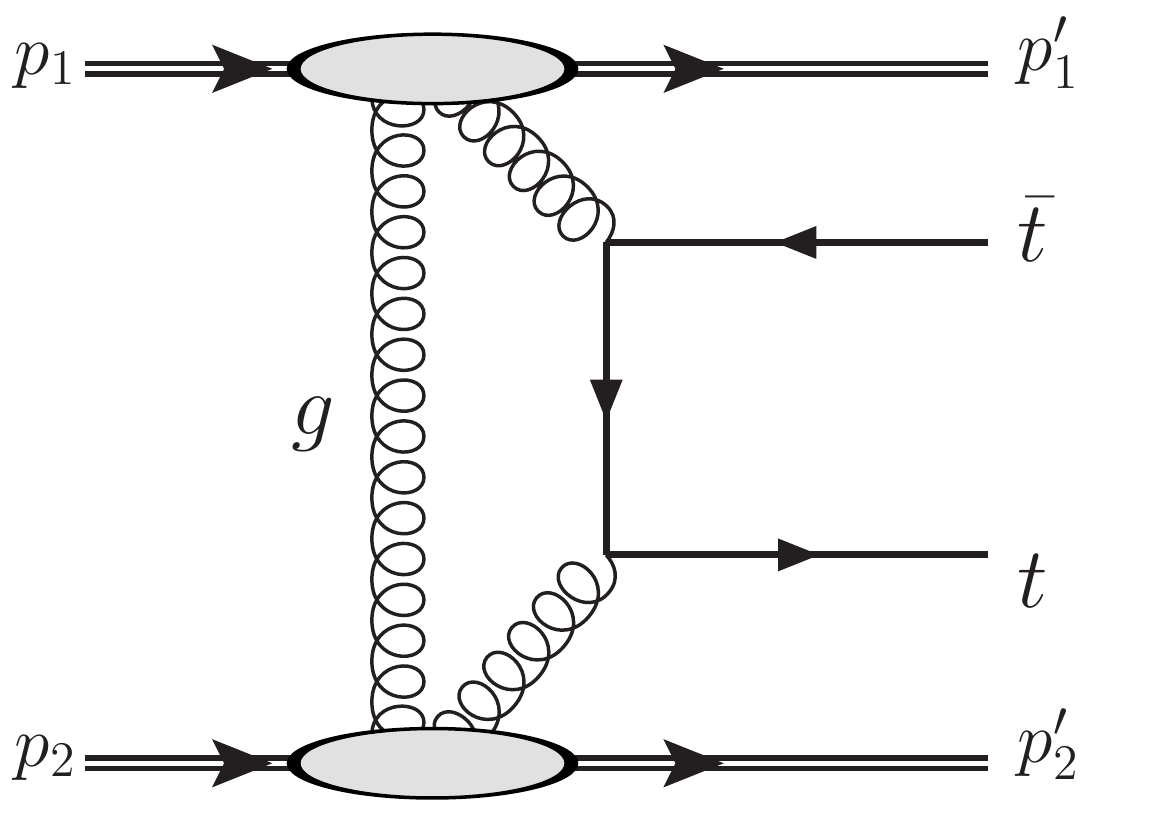}
         \caption{}
         \label{fig:SM_QCD}
     \end{subfigure}
     \begin{subfigure}[b]{0.31\textwidth}
         \centering
         \includegraphics[width=\textwidth]{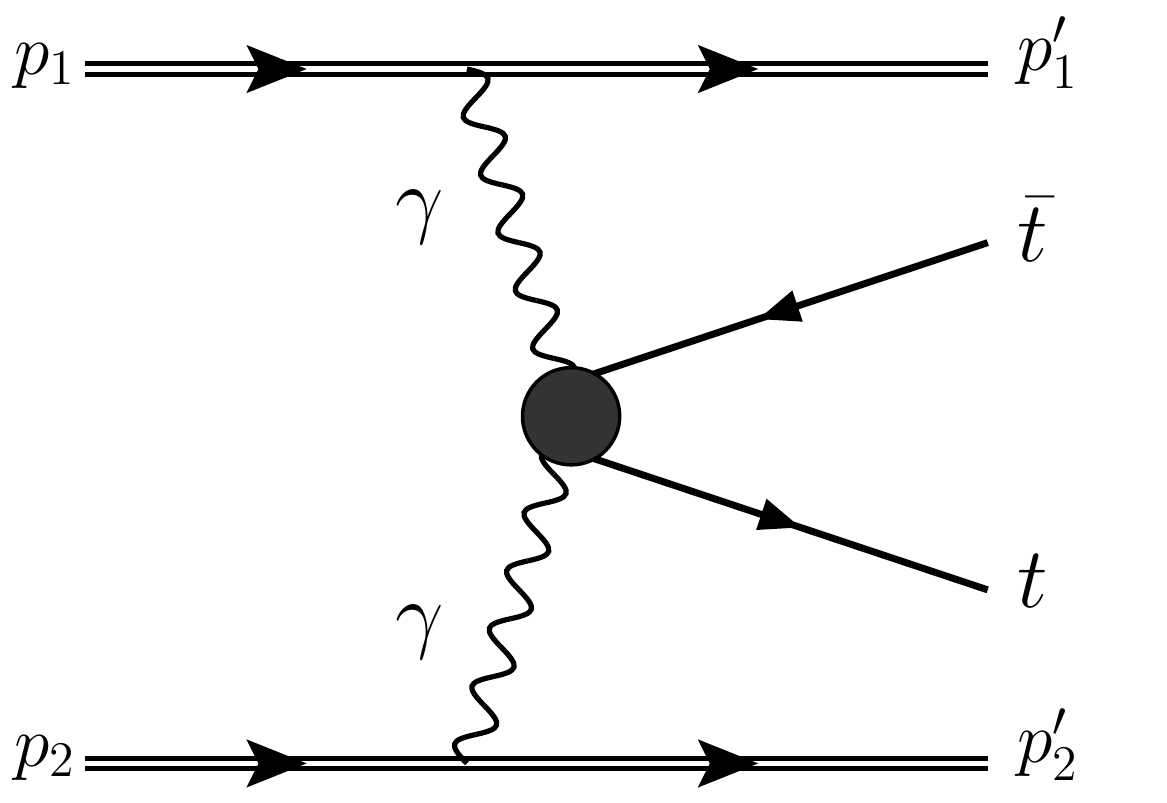}
         \caption{}
         \label{fig:BSM}
     \end{subfigure}     
     \caption{Feynman diagram for exclusive production of $t\bar{t}$ via photon fusion (a), color-singlet two-gluon fusion (b), and anomalous exclusive production via a local effective operator from the basis\ \eqref{eq:EFTbasis} (c). }
     \label{fig:diagrams}
\end{figure}

In our setup the anomalous $pp\to t\bar t pp $ process (Fig.\,\ref{fig:BSM}) is probed in the configuration of forward protons, {\it i.e.} in very peripheral proton-proton collisions. 
In very peripheral collisions, the electromagnetic field generated by the electrically charged protons can be treated as a source of quasi-real photons. These quasi-real photons can then interact to produce the top quark pair system. This photon emission off the proton is treated within the equivalent photon approximation (EPA) \cite{epa1,epa2}. In contrast to the standard parton distribution functions used in the quantum chromodynamics (QCD) factorization theorem, the flux and energy spectra of the quasi-real photons are under better theoretical control. This is because they are parametrized with the elastic electric and magnetic form factors extracted from precision electron-proton elastic scattering data. Due to additional soft interactions between the protons, not all of them will remain intact after the quasi-real photon exchange. This means that the cross section for photon-induced processes with intact protons is reduced~\cite{Khoze:2001xm,Khoze:2002dc,Gotsman:2005rt,Harland-Lang:2015cta,Harland-Lang:2018iur}. For our projections, this effect is taken into account by supplementing the cross section calculation with a survival probability factor that accounts for the probability that the protons remain intact after the photon exchange. For our calculations, we assume a survival probability of 90\%.

The signal process originates from the squared anomalous amplitude plus the interference with the SM exclusive production of $t\bar{t}$. The SM exclusive production of $t\bar{t}$ mainly happens via photon-initiated interaction (Fig.\,\ref{fig:SM_QED}), while the amplitude for gluon initiated CEP (Fig.\,\ref{fig:SM_QCD}), which can be calculated with the perturbative QCD Durham model \cite{kmr}, is at least two orders of magnitude lower, thus negligible. 

In the present study the SM cross section at $\sqrt{s}=14$\,TeV, $\sim0.35$ \cite{Goncalves:2020saa,d_Enterria_2010}, is between one and two orders of magnitude lower than its anomalous counterpart when considering coupling values equal to the predicted sensitivity. This difference is further enhanced when applying the  selection criteria. 
The fact that the square SM amplitude  is smaller than the square BSM amplitude necessarily implies that the latter dominates the signal while the interference is subleading (see \cite{Fichet:2016iuo}). Anticipating the final sensitivity, we can thus neglect beforehand the interference between anomalous and SM amplitudes. 

The anomalous central exclusive production of $t\bar{t}$ via $\gamma\gamma$ interaction (Fig.\,\ref{fig:BSM}) is generated with the Forward Physics Monte Carlo (FPMC) event generator \cite{fpmc}, which implements the EPA approach. The scattering amplitudes were computed using MadGraph \cite{mg5}, and implemented in FPMC.

\subsection{Background modeling}

The present study focuses on the reconstruction of anomalous exclusive $t\bar{t}$ production in the semi-leptonic decay channel. The dominant background components to $t\bar{t}$ exclusive production, in high-pileup conditions, are processes with final states similar to $t\bar{t}$ semi-leptonic decays, in coincidence with intact outgoing protons coming from soft diffractive pileup events. A next-to-leading order cross section of $903$ pb for QCD $t \bar t$ production at $14$ TeV with resummation at next-to-leading logarithmic accuracy is considered following Ref.~\cite{ttxsec}.

Soft diffractive reactions at $14$ TeV are expected to have a production cross section  of the order of $20$ mb. During the CERN LHC Run 2, average simultaneous interactions (pileup) in the $30$-$50$ range were reached at the ATLAS and CMS interaction points. This high number of interactions, combined with the large soft single and central diffractive cross section yields a significant amount of forward protons coming from pileup collisions. Even if kinematic constraints between forward protons and the central system measurement can be exploited to tame this background component,  as shown in the later sections, it still represents the main challenge when searching for $t\bar{t}$ exclusive production. 

To take into account this effect, a conservative average number of concurrent interactions of $50$ is considered. In each $t\bar t$ event, the number of pileup interactions is sampled from a Poisson distribution. For each pile up vertex, 0, 1 (in one detector arm) or 2 protons (in both arms) are assigned. The probability per-pileup vertex of having protons in the forward detectors is calculated using the Pythia8 event generator~\cite{pythia8}. Then, the $\xi$ distribution for these intact protons is sampled from a $1/\xi$ distribution, which is an approximation to the expected $\xi$ distribution of soft diffractive interactions. The proton fractional momentum loss is considered in the range between $0.015$ and $0.2$ for our projections. A $2\%$ Gaussian smearing is applied on $\xi$ to mimic detector resolution effects.  

The main background component is given by QCD $t\bar{t} + jets$ production. Simulated events for this process are generated with MadGraph and matched to parton showers and hadronization with Pythia8. A secondary background contribution is the QCD $WW + jets$ production in coincidence with pileup protons from soft diffraction. In this case the cross section considered is $131.3$ pb, following Ref.~\cite{wwxsec}.

Another background contribution might arise from single-diffractive $t\bar{t}$ production in coincidence with pileup protons. The process cross section is expected to be $\mathcal{O}(1)$ pb. However, even though one of the intact protons might be kinematically correlated to the central system, the second pileup proton will not. Hence, it is effectively suppressed by exploiting this kinematical correlation, making it negligible.

In double-pomeron exchange, not all of the energy of each exchanged pomeron goes into the production of the $t\bar{t}$ system. This means that the invariant mass of the $t\bar{t}$ system reconstructed with the intact protons information heavily underestimates the true $t\bar{t}$ mass (see for example the diffractive mass fraction reported by CDF in \cite{CDF:2000odl} or Figure 3 of Ref.~\cite{Martins:2022dfg}). This process is therefore suppressed in the selection process.

CEP of $t\bar{t}$ by color-singlet two-gluon exchange, which is an irreducible background component, is suppressed at high $t\bar{t}$ masses. This is because the cross section needs to be supplemented with a Sudakov form factor that accounts for the probability that there is no initial-state radiation that could spoil the exclusivity of the process. The probability that there are no gluon emissions decreases rapidly with the mass of the central system, since there is more phase-space available for them. No such Sudakov form factor is present in calculations with two-photon fusion processes. This has been shown explicitly in Refs.~\cite{Fichet:2014uka, Harland-Lang:2018iur}.

\subsection{Detector simulation}

The detector simulation is performed with the Delphes package, tuned with the CMS datacard \cite{de2014delphes}. Detector efficiencies and energy/momentum smearings are parametrized as a function of the particle kinematics (pseudo-rapidity, transverse momentum for tracking detectors, pseudo-rapidity and energy for calorimeters) \cite{cms_electron,cms_ecal}. 
Jets are clustered with the FastJet package \cite{Cacciari:2011ma}, using the anti-$k_t$ algorithm\cite{Cacciari:2008gp} with a distance parameter $R=\sqrt{\Delta \phi^2 + \Delta y^2}=0.5$, and transverse momentum $p_T > 20 $ GeV. $b$-tagging is included following Ref.~\cite{CMS_BTagging} and takes into account both the tagging efficiency and the misidentification rate. Leptons are reconstructed within the pseudo-rapidity range $|\eta| < 2.5$, and jets for $|\eta| < 5$. Missing transverse energy is computed in a particle-flow approach.

To account for the proton measurement resolution, a $2\%$ Gaussian smearing on the proton fractional momentum loss ($\xi = (p_i - p_f)/p_i$) is applied and only protons with $0.015 < \xi < 0.20$ are considered, to match the typical forward detectors acceptance. 

The longitudinal position of primary vertices is randomly sampled from a Gaussian distribution of $5$ cm RMS, according to the reported CMS observations \cite{cms_vertex}. Timing detector resolution effects are included by applying Gaussian smearings on the proton time of arrival, assuming a $200$ m distance of forward detectors from the interaction point. Timing resolutions of $20$ and $60$\,ps are considered, the former, being compatible with the design scenario presented in \cite{PPStdr}, and the latter accounting for performance already achieved today.

\section{Statistical framework}
\label{se:stats}

For our projections, we need to assume a set of observed data. As commonly done, we assume that no statistical fluctuations are present in these pseudo-data, which are usually dubbed ``Asimov'' data and that we denote here with a prime. As mentioned before, the integrated luminosity is chosen to be $L=300$\,fb$^{-1}$, which is about the expected integrated luminosity to be taken by CT-PPS and AFP in Run-3 (in Run-2, CMS and ATLAS collected about 100 of data with forward detectors at 13 TeV).

The observed events follow a Poisson distribution and for this kind of analysis we can safely neglect the systematic uncertainties. 
The statistics for the events together with the prediction for the event rates  sets the likelihood function needed for our analysis,
\be
{\cal L}(\sigma )={\rm Pr}( n' |b+\sigma L)\,,\quad {\rm Pr}( \hat n |n)=\frac{ {n^{\hat n}} e^{-n  } }{\hat n !}\, \label{eq:L}
\ee
where $b$ is the expected number of events from the background. 
In our study, depending on the cases, $b$ may be much larger than $1$ or $O(1)$. 
In order to draw projected exclusion contours, we take the background-only hypothesis for the Asimov data, \be n'=b\,. \ee

Our method to obtain the projected exclusion contours is the following. We define the posterior probability density for $\sigma$ as $L(\sigma)\pi(\sigma)$ where the prior is $\pi(\sigma)=1$ if $\sigma>0$ and $0$ otherwise.
The highest posterior density region at $1-\alpha $ credibility level is solved analytically and is simply given by
 \be
 1-\alpha=\frac{\int^{\sigma_\alpha}_0 L(\sigma) \pi(\sigma)}
 {\int_{0}^\infty L(\sigma)\pi(\sigma)}= 1-\frac{\Gamma(1+b,b+\sigma L )}{\Gamma(1+b,b)}\,, \label{eq:CL_def}
 \ee
where $\Gamma(x,y)$ is the incomplete Gamma function. 
The credibility regions for different values of   $\alpha$ amount respectively to exclusion at $2\sigma$, $3\sigma$, $5\sigma$ in terms of standard statistical significance~\footnote{
The standard statistical significance is given by $Z_\alpha=\Phi^{-1}(1-\alpha)=-\sqrt{2}{\rm Erf}^{-1}(2\alpha-1)$. We can also notice that for $b\gg 1$, upon approximating $\Gamma(x,y)\sim y^{x-1}e^{-y}$ and expanding in small $\frac{s}{b}$ we obtain $\frac{s_\alpha}{\sqrt{b}}=\sqrt{-2\log\alpha}$, where $s_\alpha=\sigma_\alpha L$. If $\alpha\ll 1$ we also have $Z_\alpha\approx \sqrt{-2\log\alpha}$, therefore in this limit our formula  reproduces the standard relation $Z_\alpha=\frac{s_\alpha}{\sqrt{b}}(1+O(\frac{s}{b}))$. 
}.
The projected sensitivity to the signal that we quote is the value of $\sigma$ satisfying  Eq.\,\eqref{eq:CL_def} for  a given statistical significance.

\section{Anomalous \texorpdfstring{$\gamma\gamma\rightarrow t\bar{t}$}{Lg} in the semi-leptonic decay channel}
\label{se:lchannel}
Since each of the top quarks decays to a $W$ boson and a $b$ quark, events can be classified depending on the $W$ boson decays. Events in which both $W$ boson decays are leptonic are called \textit{fully-leptonic}. When one $W$ boson decay is hadronic and other is leptonic the event is called \textit{semi-leptonic}. When two hadronic $W$ boson decays take place, events are called \textit{fully-hadronic}. 

Fully-leptonic decays generally present lower backgrounds, but have a low branching fraction of the order of $10\%$, and the presence of two neutrinos in the final state lowers the precision of the $t\bar{t}$ system mass measurement, effectively reducing the background rejection provided by the measurement of the leading protons. Fully-hadronic decays on the other hand have the highest branching fraction. They are however strongly impacted by QCD jets backgrounds, so they represent a challenging environment for exclusive $t\bar{t}$ measurements. For the present study, the event selection and reconstruction was therefore tuned for semi-leptonic events, as they represent a good compromise under all the above-mentioned aspects. 

Throughout this section the signal is taken to be generated from the basis of $t\bar t\gamma\gamma$ dimension-8 operators (see Eq.\,\eqref{eq:EFTbasis}).  The analysis is applied to signal events from neutral particles in the next section. 

\subsection{Event selection and reconstruction}
\label{sec:signal_selection}
Exclusive semi-leptonic $t\bar{t}$ events are selected in three steps: pre-selection, central selection and matching and eventually timing. A pre-selection is applied to select events with at least one lepton in the final state, two $b$-tagged jets, two non $b$-tagged jets, and one tagged proton per detector side. Furthermore, a missing energy transverse $E_T^{miss} > 20 $\,GeV is required. 

EFT operators yield $t\bar{t}$ mass distributions that can be grouped into two families: $\mathcal{O}_5$ and $\mathcal{O}_6$, which contain an additional derivative, produce cross sections that grow faster with $m_{t\bar{t}}$, compared to $\mathcal{O}_1$,$\mathcal{O}_2$,$\mathcal{O}_3$,$\mathcal{O}_4$. The two families are thus studied in parallel, and $\mathcal{O}_1$ and $\mathcal{O}_5$ are shown as their representatives. 

\begin{figure}[tp]
     \centering
     \includegraphics[width=0.7\textwidth]{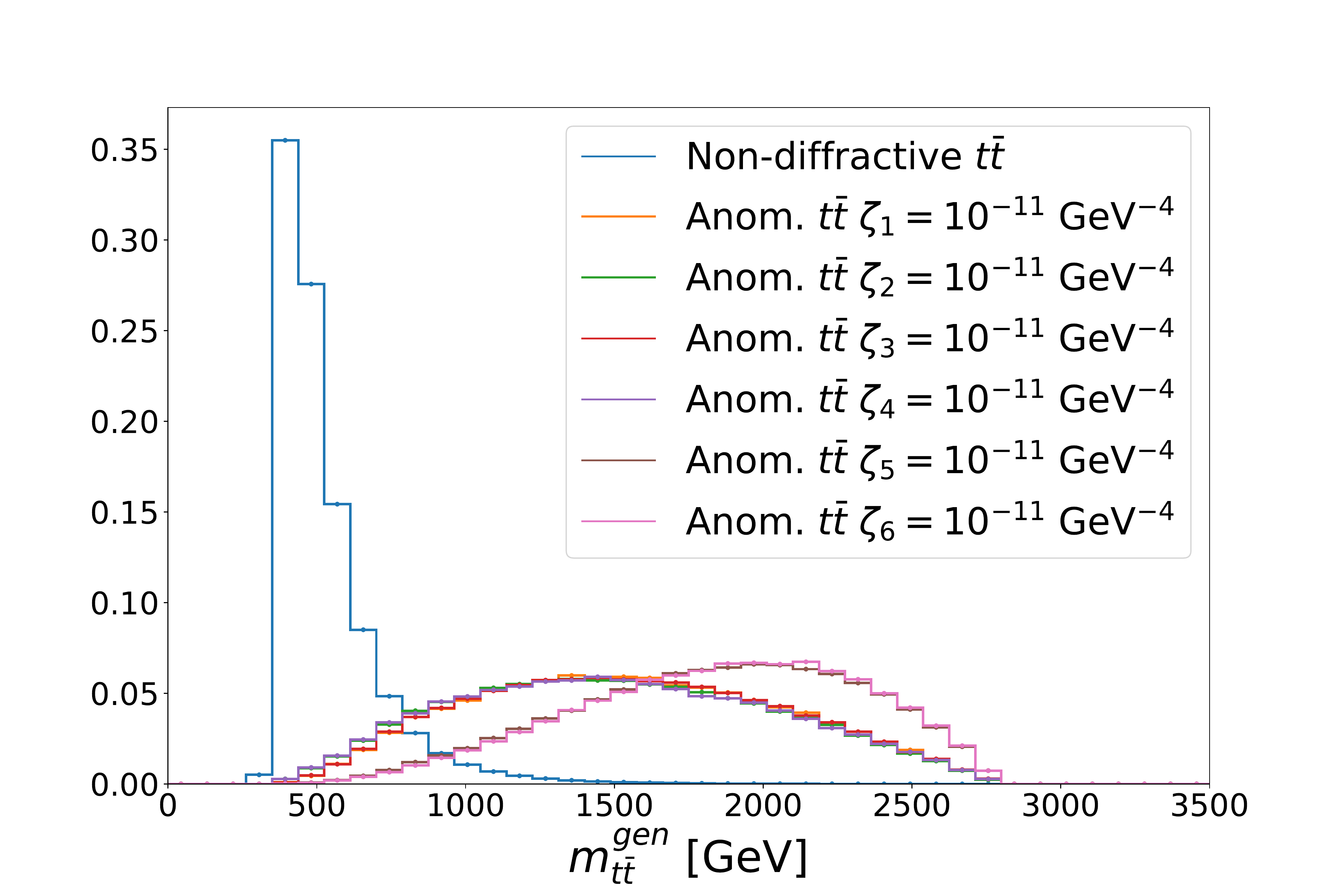}
     \caption{Mass of the $t\bar{t}$ system at generator level (stable-particle level). All histograms are normalized to unity.}
     \label{fig:mttgen}
\end{figure}

As illustrated in Fig.\,\ref{fig:mttgen}, anomalous $t\bar{t}$ events are expected to show mass distributions peaked towards higher values with respect to SM $t\bar{t}$. The two $b$-tagged jets and non $b$-tagged jets with the largest transverse momenta are therefore chosen, as well as the largest $p_T$ lepton. 

The mass of the central system can be expressed in relation to the proton measurement with:
\begin{equation}
    \label{eq:mtt_protons}
    m_{t\bar{t}}^{p} = \sqrt{s\xi_+\xi_-}
\end{equation}
with $\xi_+$ and $\xi_-$ being the fractional momentum losses of the protons measured in the two proton detectors. The proton with the largest $\xi$ is chosen to enhance the event selection efficiency of anomalous $t\bar{t}$ exclusive events in case of multiple protons on each detector side. 

After the pre-selection is performed, each top quark is reconstructed from its decay product. In the following, $t_h$ refers to the top quark that decays into hadrons, while $t_\ell$ refers to the one decaying into leptons.

To assign the $b$-tagged jets to their respective top (anti)quark, the mass of the system formed by the two non $b$-tagged jets and a $b$-tagged jet is computed. The $b$-jet assigned to $t_h$ is thus the one yielding a mass measurement closest to the top quark mass.
The other $b$-jet is consequently assigned to $t_\ell$. $t_\ell$ is reconstructed together with the highest $p_T$ lepton and missing  transverse energy. In order to improve the $t_\ell$ reconstruction, the longitudinal component of the neutrino four-momentum ($p_z^\nu$) is estimated. The value of $p_z^\nu$ is computed analytically based on the relation of the four-momentum of the lepton-neutrino system and the $W$ boson mass,
\begin{equation}
    \label{eq:pz_nu}
    \sqrt{(E^\ell + E^\nu)^2 - (\mathbf{p}^\ell + \mathbf{p}^\nu)^2} = m_W
\end{equation}
where the lepton kinematics are measured, $ p_T^\nu = E_T^{miss}$, and  the neutrino mass is neglected. 
The solution that minimizes $|p_z^\nu - p_z^\ell|$ is chosen, and in case of imaginary solutions only the real part is considered. 

\begin{figure}[tp]
     \centering
     \begin{subfigure}[b]{0.49\textwidth}
         \centering
         \includegraphics[width=\textwidth]{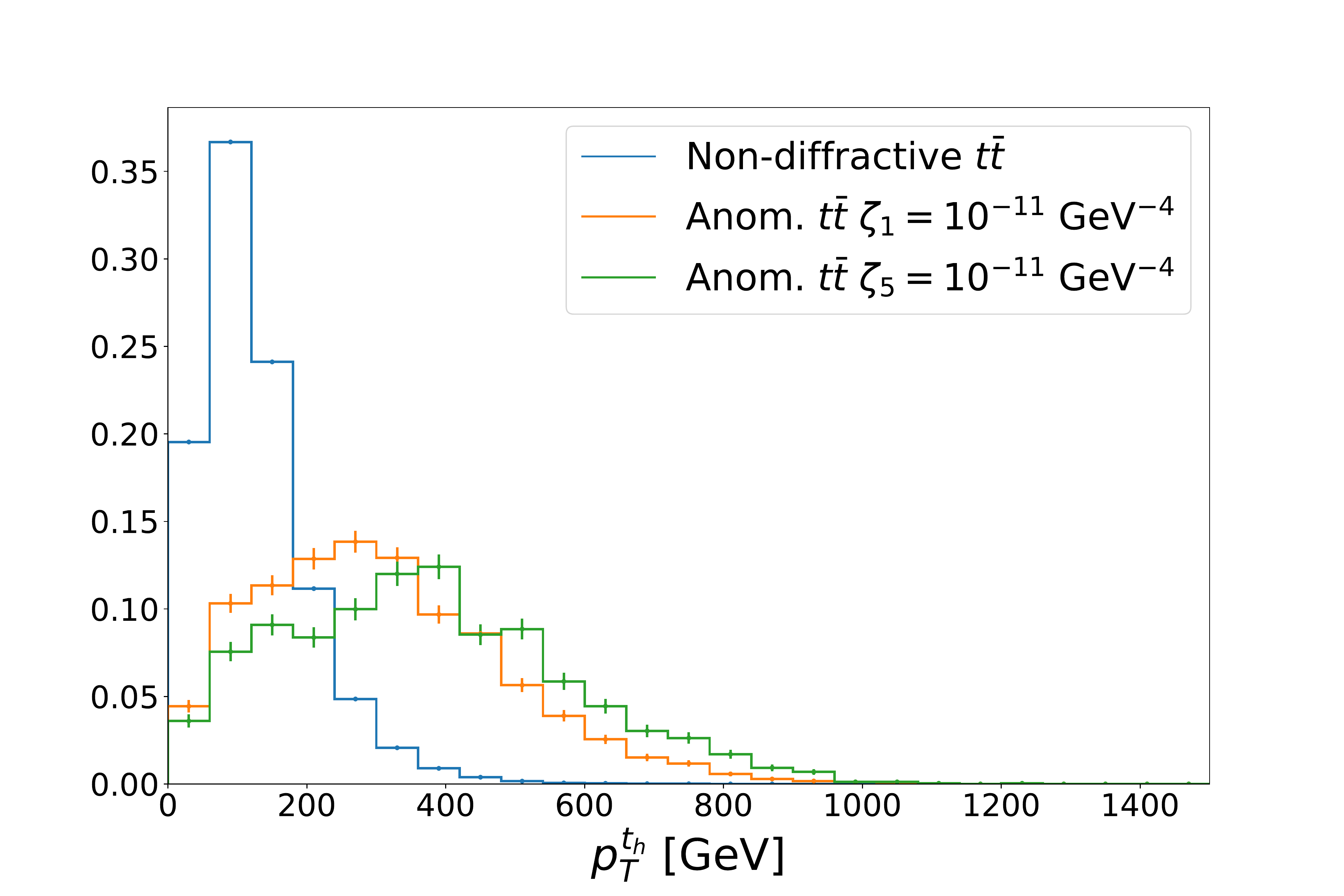}
         \caption{}
         \label{fig:pt_t1}
     \end{subfigure}
     \begin{subfigure}[b]{0.49\textwidth}
         \centering
         \includegraphics[width=\textwidth]{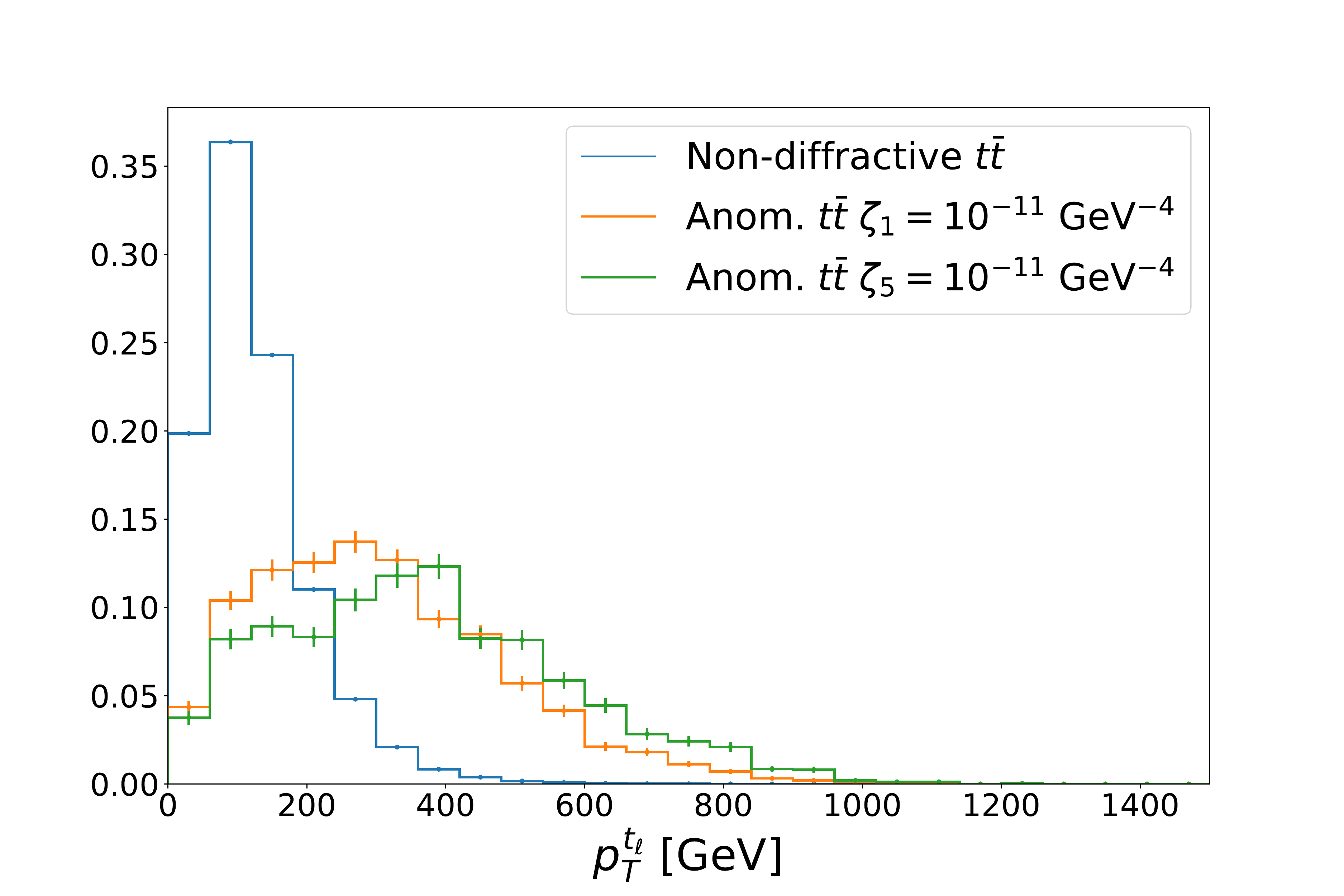}
         \caption{}
         \label{fig:pt_t2}
     \end{subfigure}
     \caption{Transverse momentum spectrum, reconstructed from $t\bar{t}$ decay products, for the top quark decaying hadronically (a) and leptonically (b). Only events passing the pre-selection are shown. Histograms are normalized to unity.}
     \label{fig:pt_t1_t2}
\end{figure}

With both $t_h$ and $t_\ell$ reconstructed, further selection is performed based on their kinematics. As shown by Fig.\,\ref{fig:pt_t1_t2}, the top quark $p_T$ spectrum in anomalous $t\bar{t}$ exclusive production is broader with respect to the SM, and higher values are more frequent. A selection of $p_T > 425$\,GeV for both $t_h$ and $t_\ell$ is therefore applied. 

\begin{figure}[tp]
     \centering
     \includegraphics[width=0.7\textwidth]{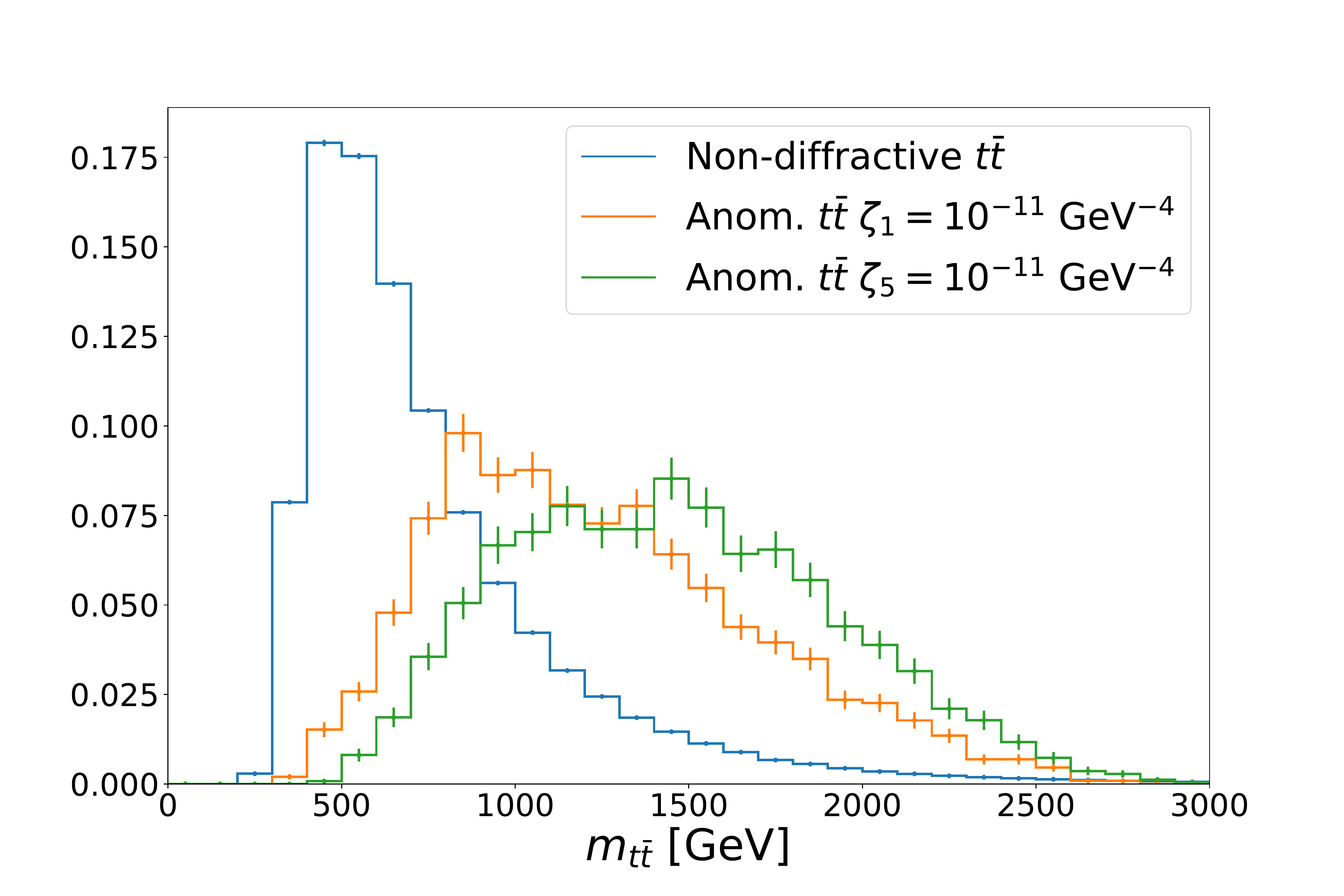}
     \caption{Invariant mass spectrum of the $t\bar{t}$ system as reconstructed from its decay products. Only events passing the pre-selection requirements are shown. Histograms are normalized to unity. The blue histogram represents the non-diffractive QCD $t\bar{t}$ component. The green and orange distributions represent the prediction of anomalous $t\bar{t}$ production assuming the coupling values of $\zeta_1 = 10^{-11}$ and  $\zeta_5 = 10^{-11}$ GeV$^{-4}$.}
     \label{fig:mtt}
\end{figure}

A similar approach is applied to $m_{t\bar{t}}$ (Fig.\,\ref{fig:mtt}), the mass measured by the central detector, selecting $m_{t\bar{t}}>960$\,GeV. A mass spectrum shifted towards higher values is indeed characteristic of anomalous exclusive processes. With this mass selection, the SM CEP of $t \bar t$ is strongly suppressed, making the interference with BSM production negligible. 

\begin{figure}[tp]
     \centering
     \begin{subfigure}[b]{0.49\textwidth}
         \centering
         \includegraphics[width=\textwidth]{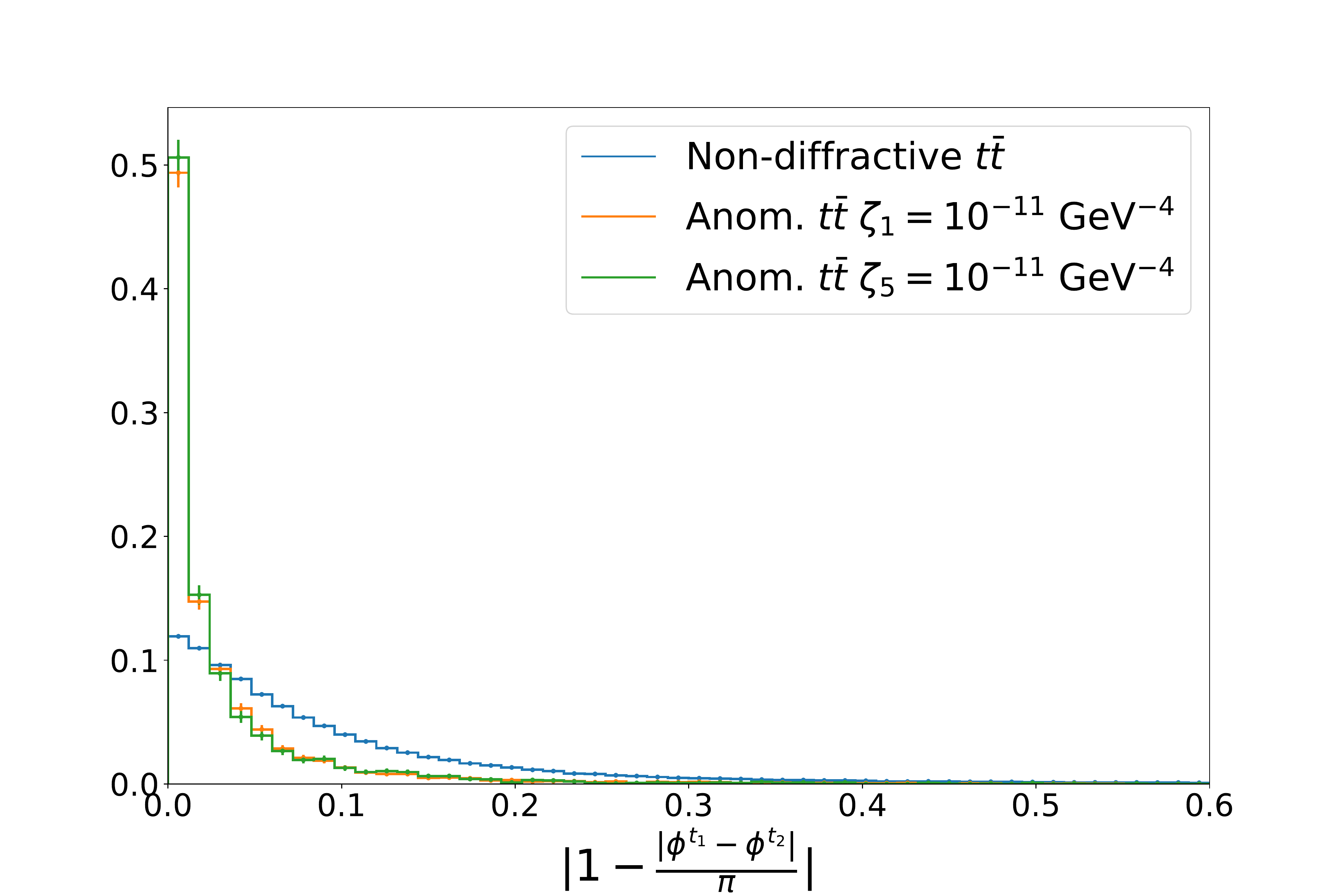}
         \caption{}
         \label{fig:acoplanarity}
     \end{subfigure}
     \begin{subfigure}[b]{0.49\textwidth}
         \centering
         \includegraphics[width=\textwidth]{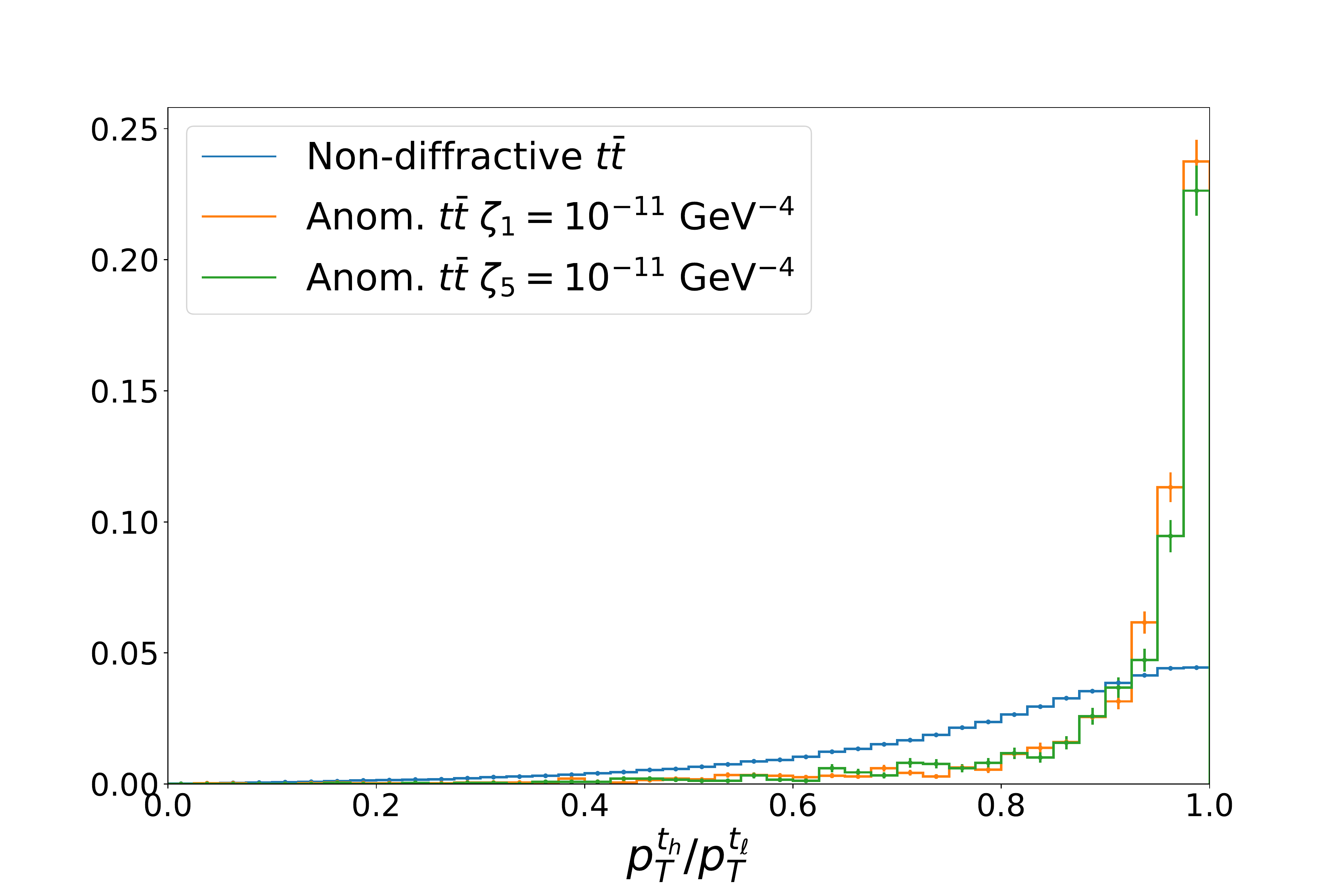}
         \caption{}
         \label{fig:pt_balance}
     \end{subfigure}
     \caption{Acoplanarity (a) and transverse momentum balance (b), reconstructed from $t\bar{t}$ decay products. Only events passing the pre-selection are shown. Histograms are normalized to unity.}
     \label{fig:acop_pt_balance}
\end{figure}

\begin{figure}[tp]
     \centering
     \includegraphics[width=0.7\textwidth]{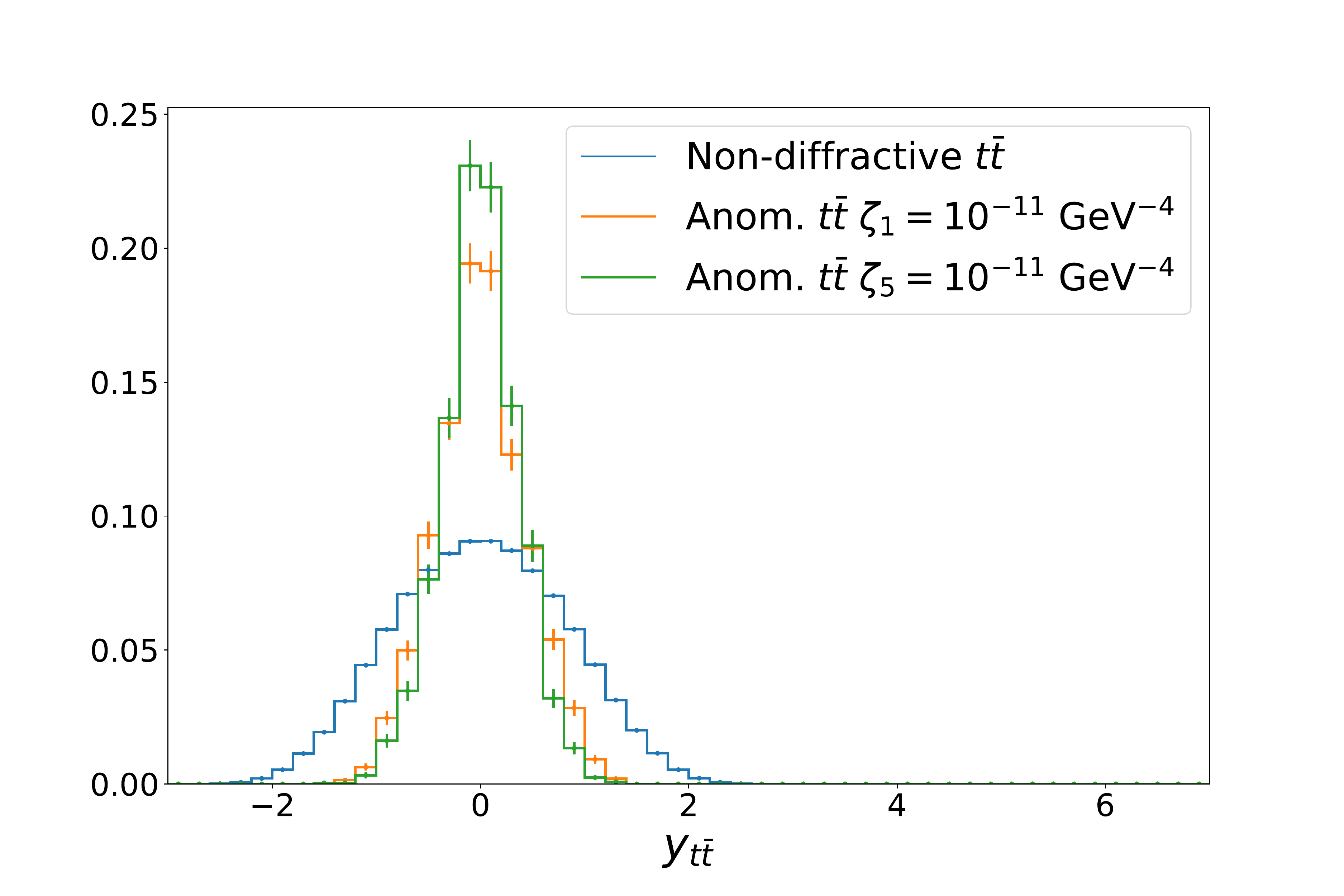}
     \caption{Rapidity distribution of the $t\bar{t}$ system as reconstructed from its decay products. Only events passing the pre-selection are shown. Histograms are normalized to unity.}
     \label{fig:ytt}
\end{figure}

Since no further radiation is expected in the final state, the tops are produced back-to-back in the transverse plane. Selecting events with acoplanarity $|1-\frac{|\phi^{t_h}-\phi^{t_\ell}|}{\pi}| < 0.09$, as Fig.\,\ref{fig:acoplanarity} illustrates, thus favors anomalous production with respect to SM $t\bar{t}+jets$ production.
Further exploiting the fact that top quarks are balanced in the transverse plane, we require the ratio of their transverse momenta $p_T^{t_h}/p_T^{t_\ell} > 0.88$ (see Fig.\,\ref{fig:pt_balance}). In exclusive events, $t\bar{t}$ pairs are also produced in a narrower rapidity range (see Fig.\,\ref{fig:ytt}), therefore we require $|y_{t\bar{t}}| < 0.72$. 

\begin{figure}[tp]
     \centering
     \includegraphics[width=0.7\textwidth]{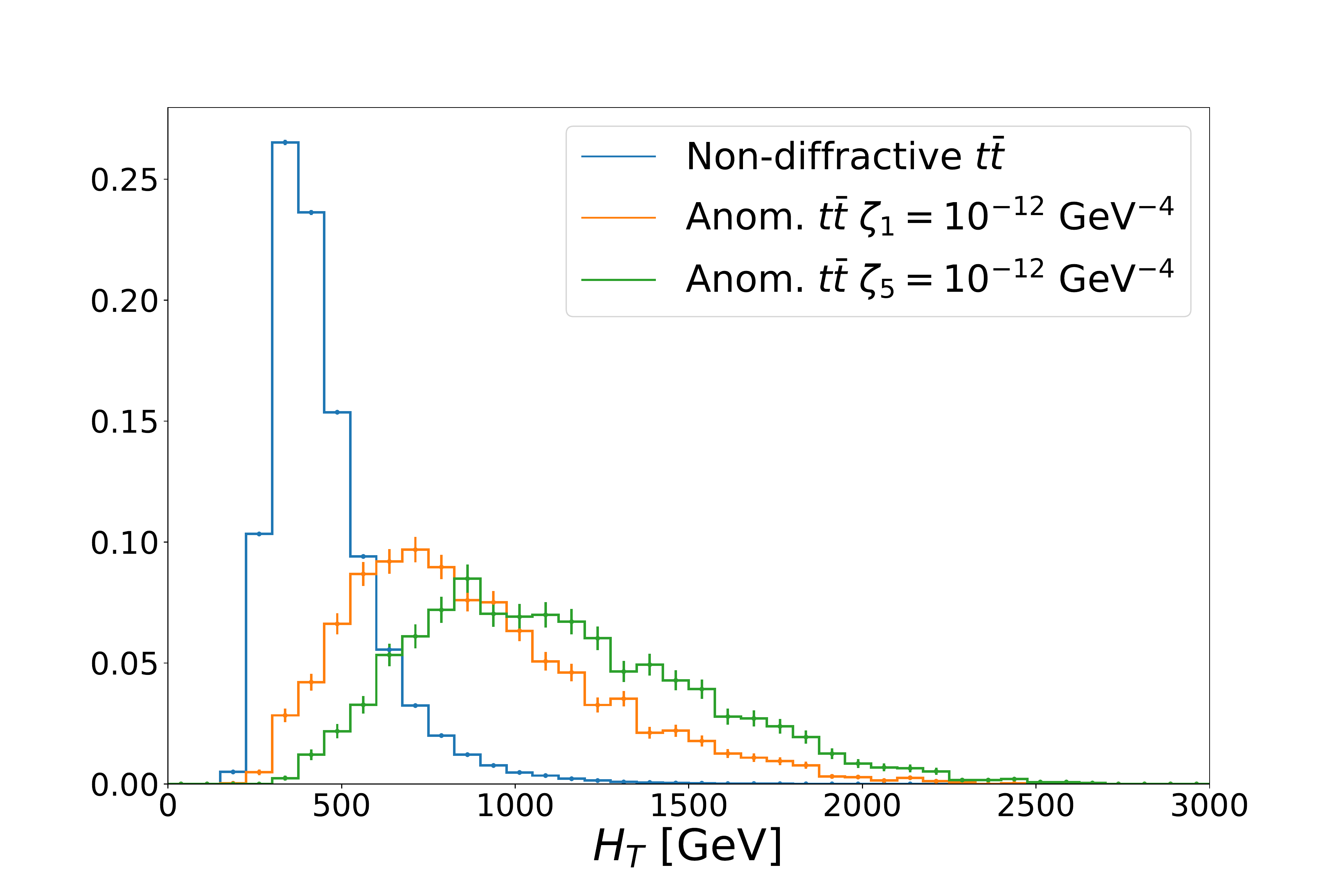}
     \caption{Sum of the transverse momenta of the decay products of the $t\bar{t}$ system ($H_T$). Only events passing the pre-selection are shown. Histograms are normalized to unity.}
     \label{fig:HT}
\end{figure}

Finally, a selection on the sum of the transverse momenta of the decay products $H_T$ is performed. $H_T$ is computed as
\begin{equation}
    \label{eq:ht}
    H_T = \sum_i p_T^i, \mathrm{with}~i=\ell,\nu,j_1,j_2,j_1^b,j_2^b
\end{equation}
where $j$ indicates the non $b$-tagged jets used for reconstruction and $j^b$ indicates the $b$-tagged ones. 
As Fig.\,\ref{fig:HT} shows, selecting high $H_T$ values above $1100$\,GeV suppresses the SM $t\bar{t}$ background while retaining anomalous exclusive $t\bar{t}$ events.

The final selection step is performed by comparing the mass and rapidity of the final state reconstructed in the central detector from the decay products with the measurements that can be obtained from proton tagging. As shown in Eq.\,\eqref{eq:mtt_protons}, the mass of the central system can be measured via the fractional momentum losses of the protons. Similarly, the rapidity of the central system can be computed from the proton kinematics:
\begin{equation}
    \label{eq:y_protons}
    y_{t\bar{t}}^{p} = \frac{1}{2} \log(\frac{\xi_+}{\xi_-})
\end{equation}
where $\xi_+$ is the fractional momentum loss of the outgoing proton with positive $p_z$, and $\xi_-$ the one of the proton with negative $p_z$.

In background events, pileup protons are uncorrelated with the processes observed in the central system.  Mass and rapidity measurements done with outgoing protons are thus expected to be uncorrelated with the same measurements given by the $t\bar{t}$ system reconstructed from its central decay products. This is reflected in variables such as the mass match ratio and $\Delta y_{t\bar{t}}$, defined as follows:
\begin{equation}
    \label{eq:massmatchratio}
    \frac{\Delta m_{t\bar{t}}}{m_{t\bar{t}}^p} = 1 - \frac{m_{t\bar{t}}^c}{m_{t\bar{t}}^p}
\end{equation}
\begin{equation}
    \label{eq:delta y}
    \Delta y_{t\bar{t}} = y_{t\bar{t}}^c - y_{t\bar{t}}^p
\end{equation}
where the $c$ index indicates reconstructed quantities from decay products measured in the central detector.

\begin{figure}[tp]
     \centering
     \begin{subfigure}[tp]{0.49\textwidth}
         \centering
         \includegraphics[width=\textwidth]{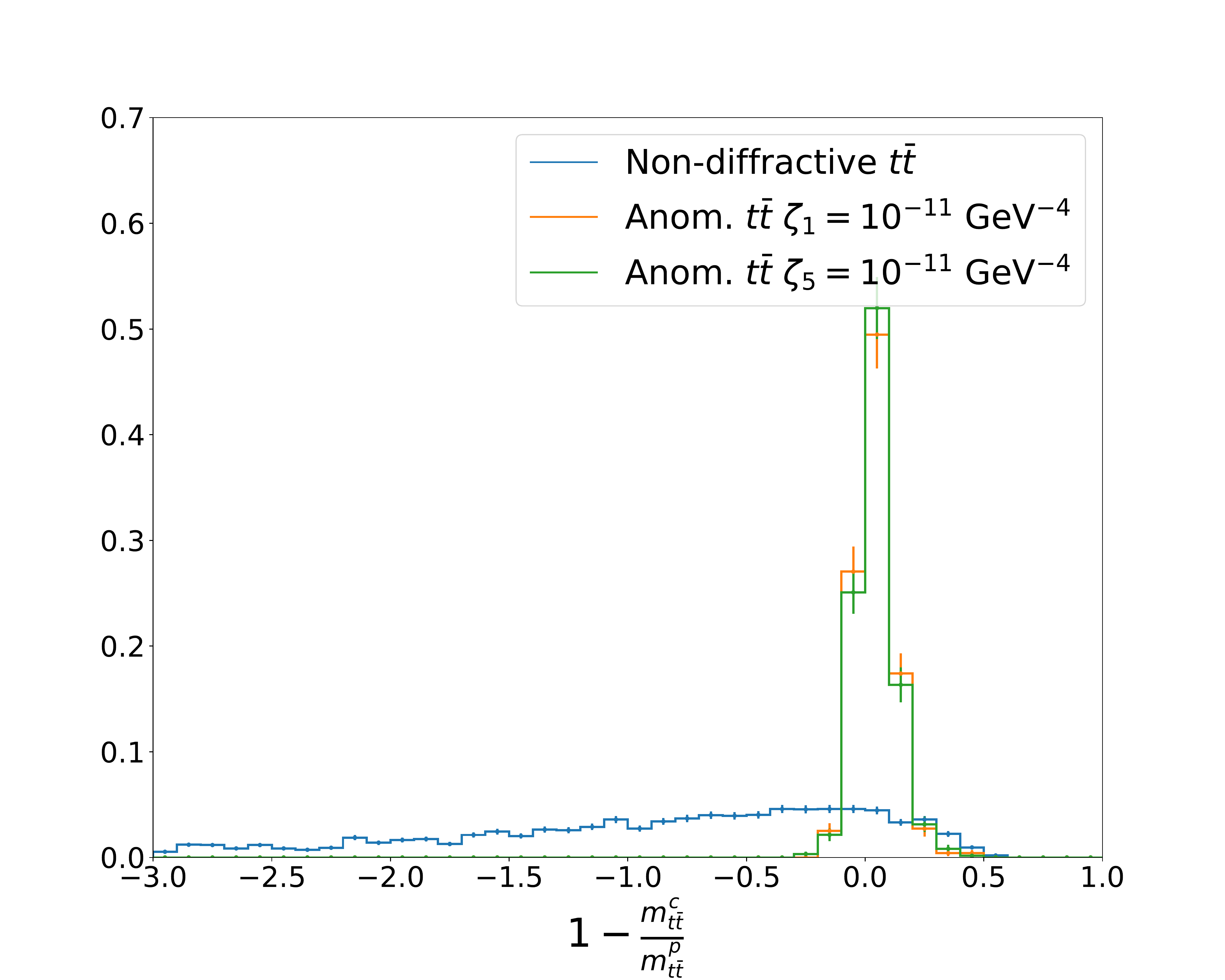}
         \caption{}
         \label{fig:massmatchratio}
     \end{subfigure}
     \begin{subfigure}[tp]{0.49\textwidth}
         \centering
         \includegraphics[width=\textwidth]{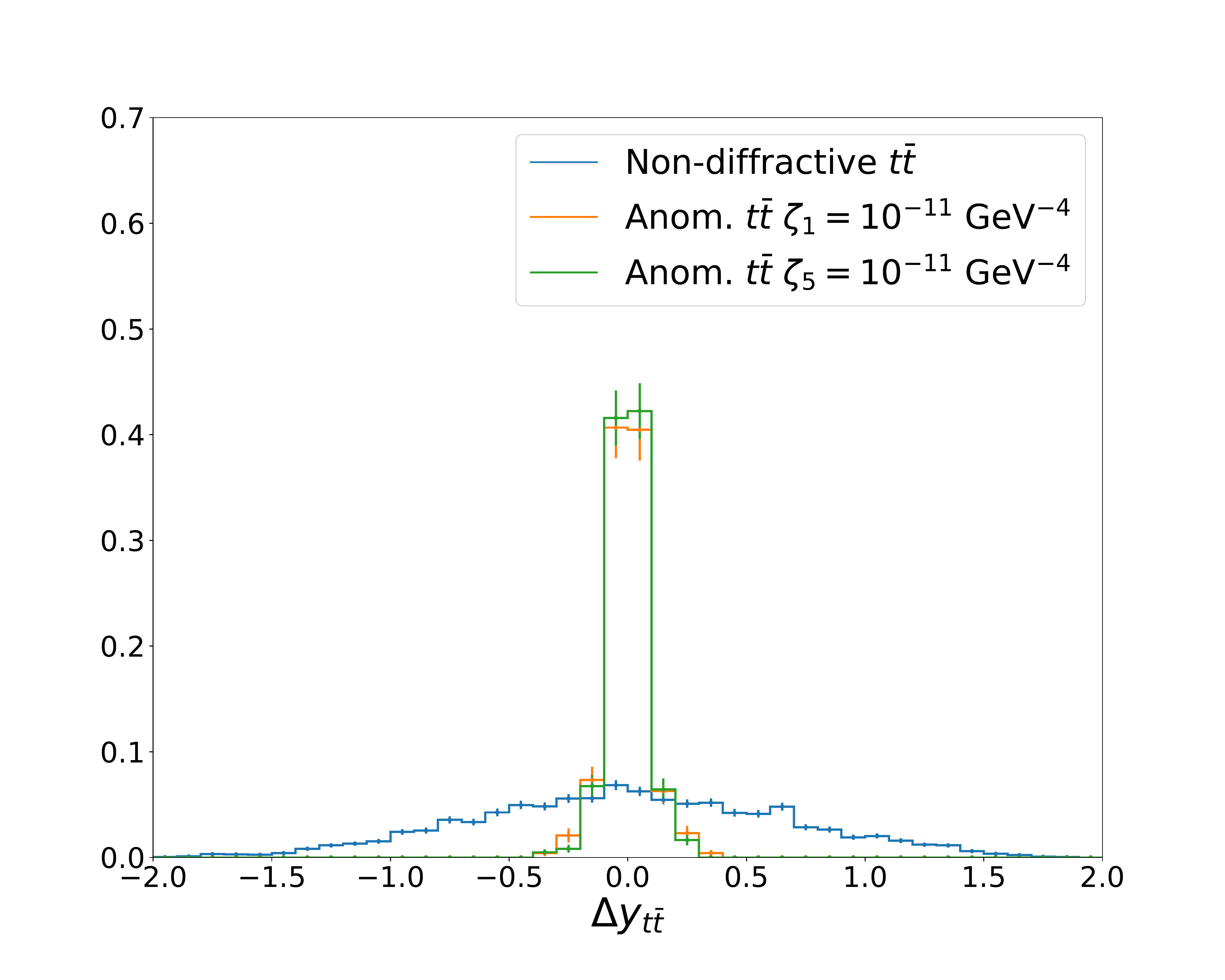}
         \caption{}
         \label{fig:deltay}
     \end{subfigure}
     \caption{Mass match ratio (a) and rapidity difference (b) between the $t\bar{t}$ reconstructed from its decay products and from tagged protons. Events passing both pre-selection and central selection are shown. Histograms are normalized to unity.}
     \label{fig:massAndRapidityMatch}
\end{figure}

\begin{table}[tp]
    \centering
    \small
    \begin{tabular}{ccccc}
\hline
 Pre-selection & Central selection & Proton matching & Timing\\
\hline
leptons $\geq 1$& $p_T^{t_h},p_T^{t_\ell} > 425$& &\\
$b$-jets $\geq 2$& $m_{t\bar{t}}>960$\,GeV& $|\frac{\Delta m_{t\bar{t}}}{m_{t\bar{t}}^p}| < 0.08$ & $|z^p-z^c| < 1.5 \sigma$\\
light jets $\geq 2$& $H_T > 1100$\,GeV& &\\
$E_\mathrm{T}^\mathrm{miss}$ $\geq 20$\,GeV& $|y_{t\bar{t}}| < 0.72$&$|\Delta y| < 0.05$&$|t_+ + t_- - \frac{2\times200 \mathrm{m}}{c}| < 1.5 \sigma$\\
protons $\geq 1$ per side& $p_T^{t_h}/p_T^{t_\ell} > 0.88$&\\
&$|1-\frac{|\phi^{t_h}-\phi^{t_\ell}|}{\pi}| < 0.09$&\\
\hline
    \end{tabular}
    \caption{Summary of the analysis sequential cuts applied at each selection stage.}
    \label{tab:summaryCuts}
\end{table}

As Fig.\,\ref{fig:massAndRapidityMatch} illustrates, signal events are expected to show a narrow distribution around zero, while background events form a broader one. This is due to the uncorrelated nature of the two pileup protons selected in background events. Since pileup protons follow a $1/\xi$ momentum loss distribution, the mass computed with a random combination peaks at lower values with respect to $m_{t\bar{t}}$, so negative mass match ratios are favored. Event candidates are classified as signal when $|\frac{\Delta m_{t\bar{t}}}{m_{t\bar{t}}^p}| < 0.08$ and $|\Delta y| < 0.05$. 

Finally, a selection based on the proton time of arrival is applied. The longitudinal vertex position reconstructed by the central tracking detectors ($z^c$) is required to be compatible with the one reconstructed from timing measurements ($z^p$), which is computed as
\begin{equation}
    \label{eq:z_p}
    z^{p} = \frac{c}{2}(t_+ - t_-)
\end{equation}
where $c$ denotes the speed of light and $t_+$ ($t_-$) indicates the time of arrival of the proton with positive (negative) $p_z$. The distribution of $z^c - z^p$ is  Gaussian in our setup, and only events falling within $1.5\sigma$ limits are accepted for maximum significance.

A similar selection is performed on the ($t_+ + t_-$) variable, which is expected to be compatible with twice the forward detectors distance from the interaction point, divided by the speed of light, in case of protons originating from the same vertex. Also in this case, a $1.5\sigma$ compatibility requirement is applied.

A summary of the cuts applied at each selection stage is given in Table\,\ref{tab:summaryCuts}.

\FloatBarrier

\section{Results}
\label{se:results}

In this section we first display our results on the $t\bar t \gamma\gamma$ dimension-8 operator basis, turning on one operator at a time. 
We then apply the analysis to models with broad neutral resonances.

\subsection{Sensitivity to dimension-8 operators}

Table~\ref{tab:signalAndBGYields} summarizes the expected events for $300 $\,fb$^{-1}$  for both signal and background, along the cut flow established in Table~\ref{tab:summaryCuts}. 

The non-diffractive $WW$ background is fully removed by the mass and rapidity matching of the central event with the proton system. 
The non-diffractive $t \bar t$ background gives $\sim 100$ events without the use of timing detectors. This is in contrast with the case of CEP of electroweak gauge boson pairs ($\gamma\gamma$, $\gamma Z$, $W^+W^-$, $ZZ$), where the non-diffractive backgrounds can be suppressed to just a few event counts\cite{ggWW_ChaponKepkaRoyon, Fichet:2014uka, Baldenegro_ALP, Baldenegro:2017aen}. The reason is that non-diffractive $t\bar{t}$ is dominated by gluon-initiated processes with a partonic cross section that scales with $\alpha_\mathrm{s}^2$, whereas non-diffractive electroweak gauge boson pair production are due to quark-initiated processes with cross sections proportional to $\alpha_\mathrm{weak}^2$ at leading order. Furthermore, the $t\bar{t}$ decay is significantly more challenging to reconstruct, which complicates the use of tighter event selection requirements to suppress the non-diffractive background. Using the timing detectors at the nominal resolutions of $60$\,ps and $20$\,ps, this background is reduced by a factor of $\sim 7$ and $\sim 60$.

We see that, if the timing detectors performance matches expectations, this channel may actually be a precision channel with almost no background, along the same lines as the well-known CEP $\gamma\gamma\to\gamma\gamma$  channel (see \textit{e.g.} \cite{Fichet:2014uka,Baldenegro_ALP}).  

The statistical significance is computed using the  expression presented in Section \ref{se:stats}, and a scan of the dimension-8 EFT operator couplings ($\zeta_{1\ldots 6}$) is performed to find the $95\%$\,CL limits and the $5\sigma$ sensitivities, presented in Table \ref{tab:resultsEFT}. The results are presented for three possible scenarios, considering a $300 $\,fb$^{-1}$ integrated luminosity: without using information from timing detectors, and using them with nominal resolutions of $60$\,ps and $20$\,ps.  
 
\begin{table}[h]
    \centering
     \footnotesize
    \begin{tabular}{l|cc|cccc}
    \hline
     & \multicolumn{2}{|c}{Signal } & \multicolumn{3}{|c}{Background} \\
     Selection step    &   $\zeta_1 = 5\cdot 10^{-11}$ &   $\zeta_5 = 5\cdot 10^{-11}$ &  $t\bar{t}$ &  $WW$ & CEP $t\bar t$ \\
        &   $\,\quad\mathrm{GeV^{-4}}$&   $\,\quad\mathrm{GeV^{-4}}$& (non-diffractive) & (non-diffractive) & (diffractive)\\
         \hline\hline
    Pre-selection     &                          $2.6\cdot10^3$ &                          $2.2\cdot10^3$ & $5.0\cdot10^6$ & $3.0\cdot10^3$ & 13\\
    Central selection &                           341 &                           487 & $5.5\cdot10^3$  & 25 & 0\\
    Proton matching   &                           246 &                           355 & 95              & 0 & 0\\
    \hline
    Timing (60\,ps)    &                           224 &                           323 & 13.8            & 0 & 0\\
    \hline
    Timing (20\,ps)    &                           224 &                           323 & 1.7             & 0 & 0\\
    \hline
    \end{tabular}
    \caption{Expected events for $300 $\,fb$^{-1}$ at each of the selection steps, for the representatives of the  operator sets ${\cal O}_{1\ldots 4 }$, ${\cal O}_{5\ldots 6 }$  and for backgrounds. All other couplings are fixed to zero. Differences among samples of the same family are within Monte Carlo statistical fluctuations.}
    \label{tab:signalAndBGYields}
\end{table}

\begin{table}[htp]
    \centering
    \footnotesize
    \begin{tabular}{ccccccc}
\hline
 Coupling $[\mathrm{10^{-11}\,GeV^{-4}}]$   &   95\% CL &   $5\sigma$ &   95\%\,CL ($60$\,ps) &   $5\sigma$ ($60$\,ps) &   95\%\,CL ($20$\,ps) &   $5\sigma$ ($20$\,ps) \\
\hline
 $\zeta_1$                               &       1.5 &        2.5 & 1.1 &        1.9 & 0.74 &        1.5 \\
 $\zeta_2$                               &       1.4 &        2.4 & 1.0 &        1.7 & 0.70 &        1.4 \\
 $\zeta_3$                               &       1.4 &        2.4 & 1.0 &        1.7 & 0.70 &        1.4 \\
 $\zeta_4$                               &       1.5 &        2.5 & 1.0 &        1.8 & 0.73 &        1.4 \\
 $\zeta_5$                               &       1.2 &        2.0 & 0.84 &        1.5 & 0.60 &        1.2 \\
 $\zeta_6$                               &       1.3 &        2.2 & 0.92 &        1.6 & 0.66 &        1.3 \\
\hline
    \end{tabular}
    \caption{ $95$\%\,CL and $5\sigma$ projected limits on each of the couplings, setting the other ones to zero.  Multiple timing detector performance scenarios are considered: no timing information, timing detector resolution  $\sigma_t =60$\,ps, and timing detector resolution  $\sigma_t =20$\,ps.}
    \label{tab:resultsEFT}
\end{table}

\FloatBarrier

\subsection{Sensitivity to  neutral scalars }

In this section we consider a  model of neutral scalar coupling to photons and top quarks (see section \ref{se:theory}). 
In such a case the direct competition with other channels should be taken into account. 
We have verified that for any value of mass and coupling, a $5\sigma$ discovery of the neutral scalar would first occur in the $\gamma\gamma\to\gamma\gamma$ channel according to the analysis from \cite{Baldenegro:2018hng}.  
As a result the question that should be raised is the following: In the hypothesis of a discovery of the neutral scalar in the $\gamma\gamma\to \gamma\gamma$ channel, how well could we measure the scalar's coupling  to top quarks?

To answer this question, we work at the level of $95\%$ confidence level. 
We assume two typical values for the coupling to the photons, 
\be
f_{\gamma\gamma}=\frac{m}{4\pi}\,,\quad \quad\quad\quad \Gamma_{\gamma\gamma} = 4\pi m \quad\quad \quad\quad{\rm (Maximally~broad~width)} 
\ee
\be
f_{\gamma\gamma}=\frac{m}{\sqrt{4\pi}}\,,\quad \quad\quad\quad \Gamma_{\gamma\gamma} =  m \quad\quad\quad\quad {\rm (Moderately~broad~width)} 
\ee
where we also indicate the value of the partial width into photons. 
For both scenarios the neutral scalar can be considered as broad since $\Gamma\geq m$. 
The case of maximally broad width also corresponds to the strongest coupling allowed by EFT validity or unitarity. 

\begin{figure}[htp]
     \centering
     \includegraphics[width=0.6\textwidth]{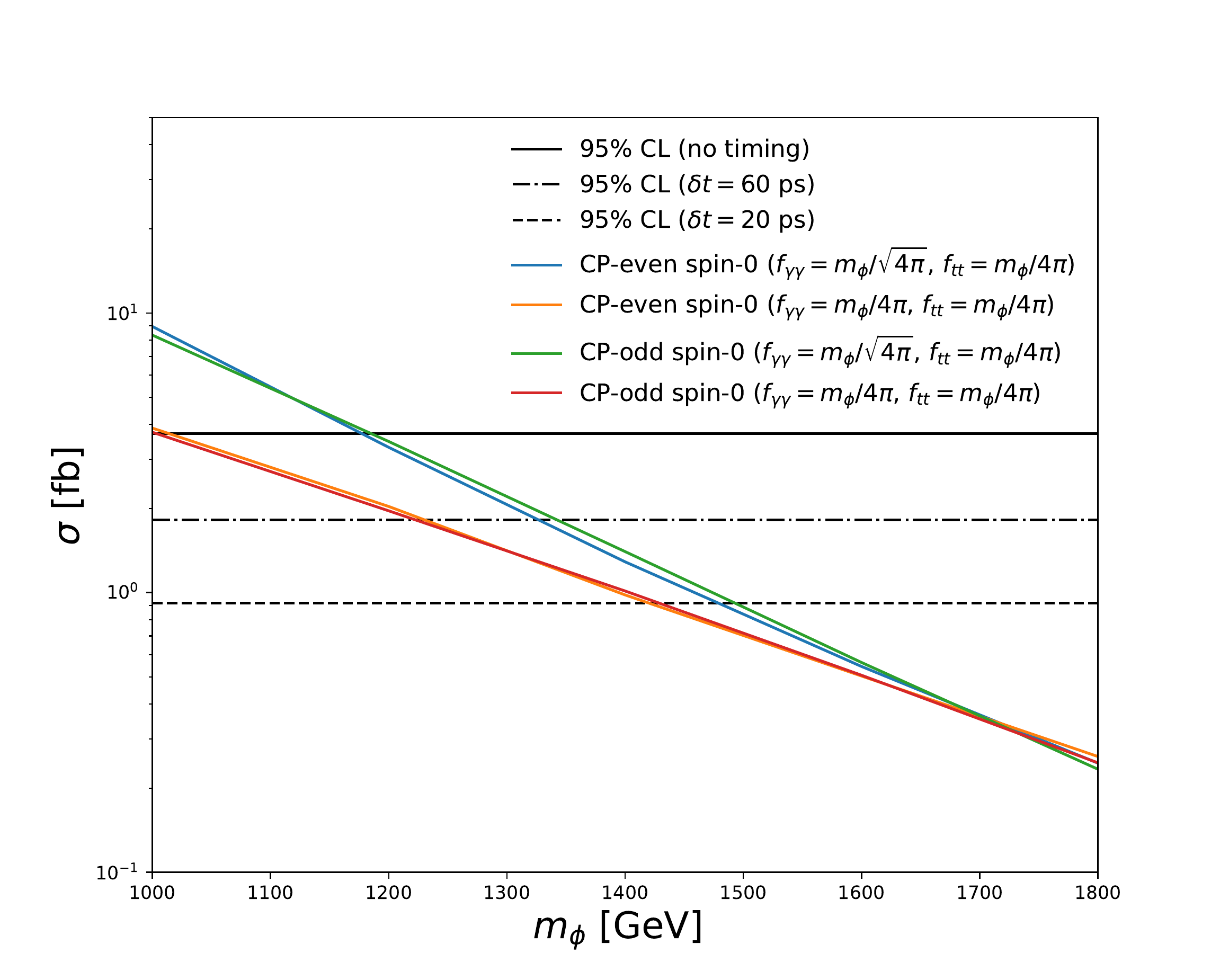}
     \caption{ Projected sensitivity to the $pp\to t\bar t\, pp $ cross section at $95\%$\,CL as a function of the mass of the neutral scalar.
     }
     \label{fig:1D}
\end{figure}

\begin{figure}[tp]
     \centering
 \includegraphics[width=0.49\textwidth]{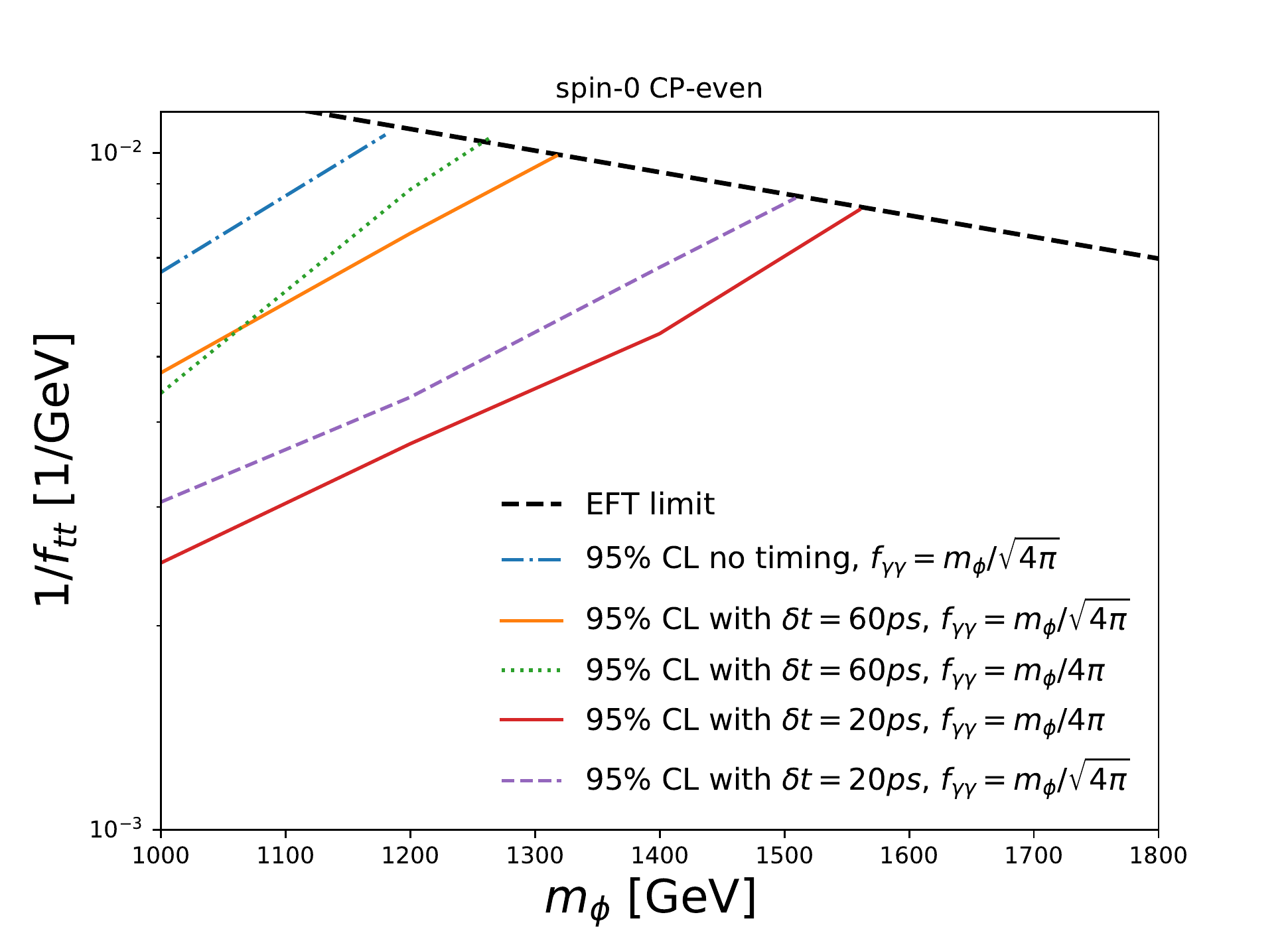}
      \includegraphics[width=0.49\textwidth]{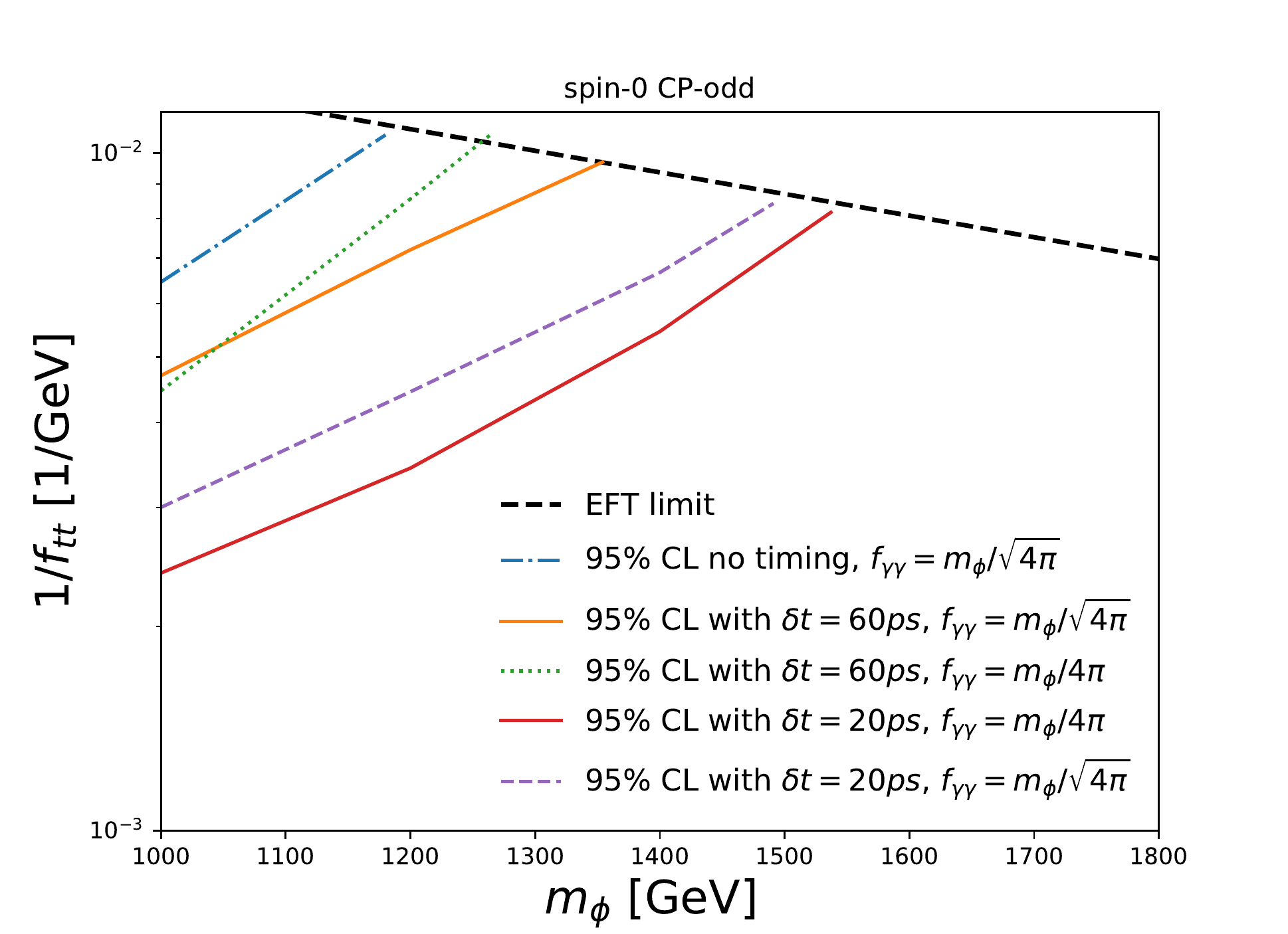}
     \caption{ Projected sensitivity at $95\%$\,CL in the $(m_\phi-f^{-1}_{\bar t t})$ plane. 
    Regions above the curves are sensitive to our $pp\to t\bar t\, pp $ search.
     }
     \label{fig:2D}
\end{figure}

In Fig.\,\ref{fig:1D} we present the projected sensitivity of our search expressed in the $pp\to t\bar t pp $ cross section. 
The CP-even and CP-odd results are very similar. 
We can see that in the mass range considered, the scenario with moderately broad width has better sensitivity. This is because the width from  $f_{\gamma\gamma}=\frac{m}{4\pi}$ is so broad that it tends to suppress the total cross section in this mass range. 

Importantly, we can see that the use of the timing detectors is key in order to increase the sensitivity of the search. Without timing detectors, the projected sensitivity in the moderately broad case goes only up to $1200$\,GeV and is vanishing for the maximally broad case over the considered mass range. In contrast, using the nominal performance for the timing detectors, it turns out that the sensitivity in both cases reaches $\sim 1450-1500$\,GeV, a substantial improvement. 
In Fig.\,\ref{fig:2D} we display the $95\%$ CL sensitivity to the $f_{t \bar t}$ coupling, either with or without timing detectors. We can again see that the timing detectors are key to expand the reach of the measurement.

\FloatBarrier

\section{Summary and perspectives}
\label{se:summary}

In this work, we have studied the sensitivity of the central exclusive $pp\to pp\,t\bar t$ process to anomalous interactions between top quarks and photons at $\sqrt{s} = 14$ TeV. The intact protons are detected with forward proton detectors --- the PPS for CMS and the AFP detector for ATLAS. We focused on the semi-leptonic decay channel of the top quark pair. We considered signals from broad neutral resonances and from $t\bar t \gamma\gamma$ dimension-8 local operators.  Our analysis does not rely on a bump search method nor on shape analysis.

The main background for central exclusive production (CEP) processes consists of quantum chromodynamics (QCD) processes paired with forward protons from uncorrelated pileup interaction. 
CEP processes are efficiently selected by using the mass and rapidity of the outgoing two-proton system provided by the forward detectors. 
Furthermore, in the channel under consideration, it turns out that the 
timing detectors complementing the forward detectors open the possibility of significant, additional background reduction.
For example, at $L=300$\,fb$^{-1}$, while the background  with no timing information is about $100$ events, the background gets reduced to $O(1)$ events when using the timing detectors with nominal resolution of $20$\,ps. This shows that the timing detectors have the potential to make the CEP $\gamma\gamma\to t\bar t$ process a high precision channel, which can be used to search for physics beyond the Standard Model. 

The projected sensitivities obtained in terms of the dimension-8 operator coefficients are of order $\sim1.5\cdot 10^{-11}$\,GeV$^{-4}$ ($95\%$ CL),
$\sim2.5 \cdot 10^{-11}$\,GeV$^{-4}$ ($5\sigma$) without timing information and are of order
$\sim 0.7 \cdot 10^{-11}$\,GeV$^{-4}$ ($95\%$CL ),
$\sim 1.5 \cdot  10^{-11}$\,GeV$^{-4}$ ($5\sigma$) when using the timing detectors with nominal resolution of $20$\,ps.

We  considered a scenario of broad neutral scalars with either moderately ($\Gamma\sim m$) or maximally ($\Gamma\sim 4\pi m$) broad width. 
We find that the timing information is key in order to significantly increase the sensitivity of the channel. For example  for resolution of $20$\,ps we find $95\%$ CL sensitivity  up to mass of $\sim 1500$ GeV, while the sensitivity is significantly worse if no timing information is used.

An extension of the present work could be to investigate hadronic decays with highly boosted  topologies. The kinematics of the new physics processes  considered in this study are such that boosted hadronic decays of top quarks can be reconstructed as large radii jets more frequently than in the SM. However, to fully exploit this channel, a judicious choice on jet substructure techniques needs to be taken to separate the boosted top quark topology from the overwhelming background of light-quark and gluon jets. For example, variables that are sensitive to the number of prongs in the large-radius jet, such as $N$-subjettiness variables, and their ratios, could be used. Also (un)groomed variables could be exploited to separate QCD top quarks, which have contributions from underlying event activity and initial-state radiation effects, from the photon-induced top quark production, which does not. We leave this as possible future work. 

The present study was only focused on dimension-8 operators, as motivated by the neutral particle scenario, which generates suppressed dimension-6 operators which have been neglected here. 
Investigating the sensitivity of CEP $t\bar t$ channels to dimension-6 photon-top quark operators (namely the top quark anomalous dipoles) is an interesting direction, especially when the timing information is taken into account. This exciting possibility is left for further investigation.

\acknowledgments

GG wishes to thank the Conselho Nacional de Desenvolvimento Cient\'ifico e Tecnol\'ogico (CNPq) for support under fellowship numbers 422227/2018-8 and 309448/2020-4 .
CB and CR thank the Brazil-US Exchange Program of the American Physical Society (APS) for the support provided in the initial stages of this study.

\bibliographystyle{JHEP}
\bibliography{references.bib}
\end{document}